%% file: eventshapes-paper-epjc.tex
\RequirePackage{fix-cm}

\documentclass[pdftex,twocolumn,epjc3]{svjour3} 

\pdfoutput=1
\RequirePackage[T1]{fontenc}
\smartqed  
\RequirePackage{graphicx}
\RequirePackage{mathptmx}      
\RequirePackage{flushend}
\RequirePackage[numbers,sort&compress]{natbib}
\RequirePackage[colorlinks,breaklinks,pdftitle={ATLAS draft},pdfauthor={The ATLAS Collaboration}]{hyperref}

\usepackage{subfigure}
\usepackage{amsmath}
\usepackage{amssymb}
\usepackage{multirow}
\usepackage{lineno}
\usepackage{hyperref}
\usepackage{xspace}
\hypersetup{
  colorlinks=true,
  linkcolor=blue,
  citecolor=blue,
  urlcolor=blue
}

\input{definitions.sty}

\graphicspath{{./}}
\usepackage{preprintcover}  
\PreprintCoverPaperTitle{Measurement of event shapes at large momentum transfer with the ATLAS detector in $\mathbf{pp}$ collisions at $\mathbf{\sqrt{s}=7}$~TeV}
\PreprintIdNumber{CERN-PH-EP-2012-119}
\PreprintCoverAbstract{A measurement of event shape variables is presented for large momentum transfer proton-proton collisions using the ATLAS detector at the Large Hadron Collider. Six event shape variables calculated using hadronic jets are studied in inclusive multi-jet events in 35\invpb of integrated luminosity at a center-of-mass energy of \sqsseven. These measurements are compared to predictions by three Monte Carlo event generators containing leading-logarithmic parton showers matched to leading order matrix elements for $2\ra2$ and $2\ra n$ ($n=2,...6$) scattering. Measurements of the third-jet resolution parameter, aplanarity, thrust, sphericity, and  transverse sphericity are generally well described. The mean value of each event shape variable is evaluated as a function of the average momentum of the two leading jets $p_{T,1}$ and $p_{T,2}$, with a mean \pT\ approaching 1~TeV.}

\PreprintJournalName{Eur.\ Phys.\ J.\ C}

\newcommand{\ra}      {\ensuremath{\rightarrow}}
\newcommand{\pt}      {\ensuremath{p_{\mathrm{T}}}}
\newcommand{\pT}      {\pt}
\newcommand{\alphas}  {\ensuremath{\alpha_{\mathrm{S}}}}

\journalname{Eur. Phys. J. C}

\begin{document}


\date{Received: date / Accepted: date}

\title{Measurement of event shapes at large momentum transfer with the ATLAS detector in $\mathbf{pp}$ collisions at $\mathbf{\sqrt{s}=7}$~TeV}

\author{The ATLAS Collaboration\thanksref{e1,addr1}}
\thankstext{e1}{e-mail: atlas.publications@cern.ch}
\institute{CERN, 1211 Geneva 23, Switzerland \label{addr1}}
                                                                                                                                                

\maketitle

\begin{abstract}

A measurement of event shape variables is presented for large momentum transfer proton-proton collisions using the ATLAS detector at the Large Hadron Collider. Six event shape variables calculated using hadronic jets are studied in inclusive multi-jet events in 35\invpb of integrated luminosity at a center-of-mass energy of \sqsseven. These measurements are compared to predictions by three Monte Carlo event generators containing leading-logarithmic parton showers matched to leading order matrix elements for $2\ra2$ and $2\ra n$ ($n=2,...6$) scattering. Measurements of the third-jet resolution parameter, aplanarity, thrust, sphericity, and  transverse sphericity are generally well described. The mean value of each event shape variable is evaluated as a function of the average momentum of the two leading jets $p_{T,1}$ and $p_{T,2}$, with a mean \pT\ approaching 1~TeV.


\end{abstract}



\section{Introduction}
\label{sec:intro}

\input{sec-intro.tex}

\section{The ATLAS Detector}
\label{sec:detector}
\input{sec-detector.tex}

\section{Data Samples and Event Selection}
\label{sec:data-selection}
\input{sec-data-selection.tex}

\section{Results and Systematic Uncertainties}
\label{sec:results}
\input{sec-results.tex}

\section{Summary}
\label{sec:conclusions}
\input{sec-conclusions.tex}

\section{Acknowledgments}
\label{sec:acknowledgments}
\input{sec-acknowledgments.tex}


\bibliographystyle{atlasnote}
\bibliography{eventshapes}
\clearpage

\clearpage
\onecolumn
\input{atlas_authlist}

\end{document}

%% file: sec-intro.tex
Event shapes represent a generic class of observables that describe the patterns, correlations, and origins of the energy flow in an interaction. In terms of hadronic jet production, event shapes are an indirect probe of multi-jet topologies. These observables have had a long and fruitful history, having been used to measure the strong coupling constant \alphas\ and to test asymptotic freedom~\cite{OpalEventShapes2005, H1EventShapes2006, ZeusEventShapes2007, JadeEventShapes2009}, to constrain color factors for quark and gluon couplings~\cite{Kluth2001}, to assess the accuracy of leading order (LO) and next-to-leading order (NLO) Monte Carlo (MC) generators~\cite{Stewart2010, Collaboration2011g}, to determine the contribution of non-perturbative quantum chromodynamics (QCD) power corrections~\cite{Webber:1998um}, and to search for physics beyond the Standard Model~\cite{Anchordoqui2010}. Furthermore, recent efforts to provide advanced, high-precision theoretical calculations of a range of event shapes for the Large Hadron Collider~\cite{Banfi2004, Banfi2010} provide renewed impetus for making such measurements.

This analysis considers six event shapes calculated using hadronic jets. These observables are crucially tied to both the multi-jet nature of the final state produced in high energy collisions and have a strong history in the literature: the third-jet resolution parameter~\cite{Akrawy1990389, Bartel1982, JadeEventShapes2009, Kluth2009}, \ytwoth; the sphericity and transverse sphericity~\cite{Bjorken1970, Barate19981}, \sph and \sphtranv; the aplanarity, \apla; and the event thrust and its minor component~\cite{Brandt1964}, \thrmaj and \thrmin. Events with high transverse momentum central leading-jet pairs are used for the measurements. Each event shape variable is defined such that it vanishes in the limit of a pure \TwoToTwo process and increases to a maximum for uniformly distributed energy within a multi-jet event. Hard gluon emission is thereby signified by large non-zero values of each observable. Furthermore, some of the event shape variables are evaluated as ratios of final state observables, which reduces their sensitivity to jet energy scale (JES) calibration uncertainties as well as other experimental and theoretical uncertainties. These measurements permit detailed tests of the phenomenological models of QCD in leading order MC programs and to indirectly test the running of \alphas\ through measurements performed as a function of the average leading jet momentum. In addition, these results may be used to provide input to tune MC generators in the future.

All event shapes measured in this analysis are defined using jets to represent the final state four-momenta, as discussed in Section~\ref{sec:defs}. The ATLAS detector is described in Section~\ref{sec:detector}, with a particular emphasis on the components relevant for the measurement of event shape variables. Section~\ref{sec:data-selection} presents the event selection and description of simulated events which are compared to the data. Jet definitions, calibrations, and selection criteria are also described in Section~\ref{sec:data-selection}. Finally, the results of these measurements are presented in Section~\ref{sec:results}.

\section{Event Shape Definitions \label{sec:defs}}

Six event shapes are measured using high transverse momentum (\pT) jets. The first observable, \ytwoth, is a measure of the third-jet \pT\ relative to the summed transverse momenta of the two leading jets in a multi-jet event and is defined as:
\begin{eqnarray}
  \ytwoth &=& \frac{p^2_{{\rm T},3}}{\HTsubTwo^{2}}, 
  \label{eq:y23}
\end{eqnarray}
where $\HTsubTwo = (\pTone+\pTtwo)$ is the scalar sum of jet momenta and the subscript $i=1,2,3$ refers to the leading, subleading, or third leading jet in the event. The range of allowed values for \ytwoth is $0\leq\ytwoth<1/4$ and it is often expressed as \lny~\cite{Barate19981, Banfi2010}. This definition is different from the original~\cite{Akrawy1990389} definition which uses the JADE jet algorithm~\cite{Bartel1986}. Eq.~\ref{eq:y23} is defined with an explicit third-jet as opposed to a continuously variable threshold in the jet algorithm. The sphericity, \sph, transverse sphericity, \sphtranv, and aplanarity, \apla, embody more global information about the full momentum tensor of the event, \Mtensor, via its eigenvalues $\lambda_1$, $\lambda_2$ and $\lambda_3$:
\begin{equation}
  \label{eq:theory:trans-spheri-matrix}
  \Mtensor = \sum_i \left( 
    \begin{array}{ccc}
      p_{xi}^2       & p_{xi} p_{yi} & p_{xi} p_{zi}\\
      p_{yi} p_{xi}  & p_{yi}^2      & p_{yi} p_{zi} \\
      p_{zi} p_{xi}  & p_{zi} p_{yi} & p_{zi}^2
    \end{array}
  \right),
\end{equation}
where the sum runs over all jets used in the measurement. The individual eigenvalues are normalized and ordered such that $\lambda_1 > \lambda_2 > \lambda_3$ and $\sum_{i}\lambda_{i}=1$ by definition. These terms are used to define the three observables as
\begin{eqnarray}
  \sph      &=& \frac{3}{2}(\lambda_2 + \lambda_3), \label{eq:theory:sphDef}\\
  \sphtranv &=& \frac{2\lambda_2}{\lambda_1 + \lambda_2}, \label{eq:theory:sphTDef}\\
  \apla     &=& \frac{3}{2}\lambda_3. \label{eq:theory:aplDef}
\end{eqnarray}
Sphericity, Eq.~(\ref{eq:theory:sphDef}), and transverse sphericity, Eq.~(\ref{eq:theory:sphTDef}), measure the total transverse momentum with respect to the sphericity axis defined by the four-momenta used for the event shape measurement (specifically, the first eigenvector). The  allowed range of \sph values is $0 \leq \sph < 1$, but due to the inclusion of the smallest eigenvalue, $\lambda_3$, the typical maximum achieved experimentally is $\sph\sim0.8$. Conversely, the transverse sphericity is constructed using the two largest eigenvalues, and the typical range coincides with the allowed range, $0 \leq \sphtranv < 1$. Aplanarity (Eq.~\ref{eq:theory:aplDef}) measures the amount of transverse momentum in or out of the plane formed by the two leading jets via only the smallest eigenvalue of \Mtensor, $\lambda_3$, with allowed values $0\leq\apla<1/2$. Typical measured values lie between $0\leq\apla<0.3$, with values near zero indicating relatively planar events. The transverse thrust, $T_{\perp}$, and its minor component, \thrmin, define a so-called thrust axis for the event, with respect to which, the total transverse momentum of the jets used in the measurement is minimized. These quantities are defined as
\begin{eqnarray}
  \label{eq:theory:thrust}
  T_{\perp} &=& \max_{\thrvec} \frac{\sum_i |{\vec p}_{{\rm T} i}\cdot {\thrvec}|}{\sum_i p_{{\rm T} i}}, \\
  \thrmaj   &=& 1 - T_{\perp}, \\
  \thrmin   &=& \frac{\sum_i |{\vec p}_{{\rm T} i}\times {\thrvec}|}{\sum_i p_{{\rm T} i}},
\end{eqnarray}
where $T_{\perp}$ is translated into \thrmaj in order to maintain a common event shape definition in which a large value indicates a departure from a two-body system. The unit vector \thrvec defines the thrust axis of the event. The so-called event plane is defined by \thrvec and the beam direction and allows a measurement of \thrmin. The variable \thrmin  quantifies the sum of all transverse momenta ${\vec p}_{{\rm T} i}$ out of the event plane, where the sum again runs over each jet $i$ considered in the final state. The allowed values for \thrmaj span the range $0\leq \thrmaj<1/3$ due to the range over which both $T_{\perp}$ and \thrmin may fall, $0\leq T_{\perp}$, $\thrmin<2/3$.

Event shapes constructed using hadronic jets in this way offer several advantages over explicit cross-section calculations for inclusive and multi-jet production. Event shapes may be defined as normalized ratios of hadronic final state observables, thus reducing the sensitivity to experimental uncertainties. Various choices of event shape quantities can also lead to enhanced or suppressed sensitivity to different components of the fundamental physical processes involved~\cite{Banfi2010}. The effect of the underlying event and parton shower can be reduced by focusing only on the leading jets. The choice of renormalization and factorization scales used in calculating the LO and NLO cross-sections may be less important when considering ratios of quantities. Systematic uncertainties, such as the jet energy scale and detector effects, are mitigated by examining the normalized shapes as opposed to absolute cross-sections.

%% file: sec-detector.tex
The ATLAS detector~\cite{detPaper, PerfWithData2010} provides nearly full solid angle coverage around the collision point\footnote{ATLAS uses a right-handed coordinate system with its origin at the nominal interaction point (IP) in the centre of the detector and the $z$-axis along the beam pipe. The $x$-axis points from the IP to the centre of the LHC ring, and the $y$ axis points upward. Cylindrical coordinates $(r,\phi)$ are used in the transverse plane, $\phi$ being the azimuthal angle around the beam pipe. The pseudorapidity is defined in terms of the polar angle $\theta$ as $\eta=-\ln\tan(\theta/2)$.} with an inner tracking system covering $|\eta|<2.5$, electromagnetic and hadronic calorimeters covering $|\eta|<4.9$, and a muon spectrometer covering $|\eta|<2.7$. Of the multiple ATLAS subsystems, the most relevant to this analysis are the inner tracking detector (ID)~\cite{InDetPerfPaper}, the barrel and end-cap calorimeters~\cite{LArReadiness, TileReadiness} and the trigger~\cite{TriggerPerf2010}. 

The ID is comprised of a pixel tracker closest to the beamline, a microstrip silicon tracker, and lastly a straw-tube transition radiation tracker at the largest radii. These systems are layered radially upon each other in the central region. A thin solenoid surrounding the tracker provides an axial 2T field enabling measurement of charged particle momenta. 

The calorimeter is built of multiple sub-detectors with several different designs spanning the pseudorapidity range up to $|\eta|<4.9$. The measurements of event shapes are predominantly performed using data from the central calorimeters, comprised of the liquid argon LAr barrel electromagnetic calorimeter ($|\eta|<1.475$) and the Tile hadronic calorimeter ($|\eta|<1.7$). Three additional calorimeter subsystems are located in the forward regions of the detector: the LAr electromagnetic end-cap calorimeters, the LAr hadronic end-cap calorimeter, and the forward calorimeter comprised of separate electromagnetic and hadronic components. 

The precision and accuracy of energy measurements made by the calorimeter system is integral to this analysis and the procedures to establish such measurements are described in Ref.~\cite{ATLASJESpaper2010}. The baseline electromagnetic (EM) energy scale of the calorimeters derives from the calibration of the electronics signal for the energy deposited by electromagnetic showers. The hadronic calorimeter has been calibrated with electrons and muons in beam tests and the energy scale has been validated using muons produced by cosmic rays with the detector {\em in situ} in the experimental hall~\cite{TileReadiness}. The invariant mass of the $Z$ boson in $Z\rightarrow ee$ events measured {\em in situ} in the same data-taking period is used to adjust the calibration for the EM calorimeters.

Dedicated trigger and data acquisition systems are responsible for the online event selection which is performed in three stages: Level 1, Level 2, and the Event Filter. Level 1 utilizes information from the calorimeter and muon systems using hardware-based algorithms. Level 2 and the Event Filter  are collectively referred to as the High Level Trigger and utilize software algorithms running on large farms of commercial processors.  The measurements presented in this paper rely primarily on the hardware-based Level 1 calorimeter trigger. At this level, coarse calorimeter information is used to reconstruct jets in the trigger system with a square sliding-window algorithm in $\eta-\phi$ space.

%% file: sec-data-selection.tex
\subsection{Data Sample and Event Selection}

The data used for the analysis of event shapes represent the entire 2010 dataset collected at \sqsseven and correspond to an integrated luminosity of $\int \lumi {\rm d}t = 35.0\pm1.1$\invpb~\cite{Collaboration2011}.

\begin{figure}
  \centering
    \includegraphics[width=0.75\columnwidth]{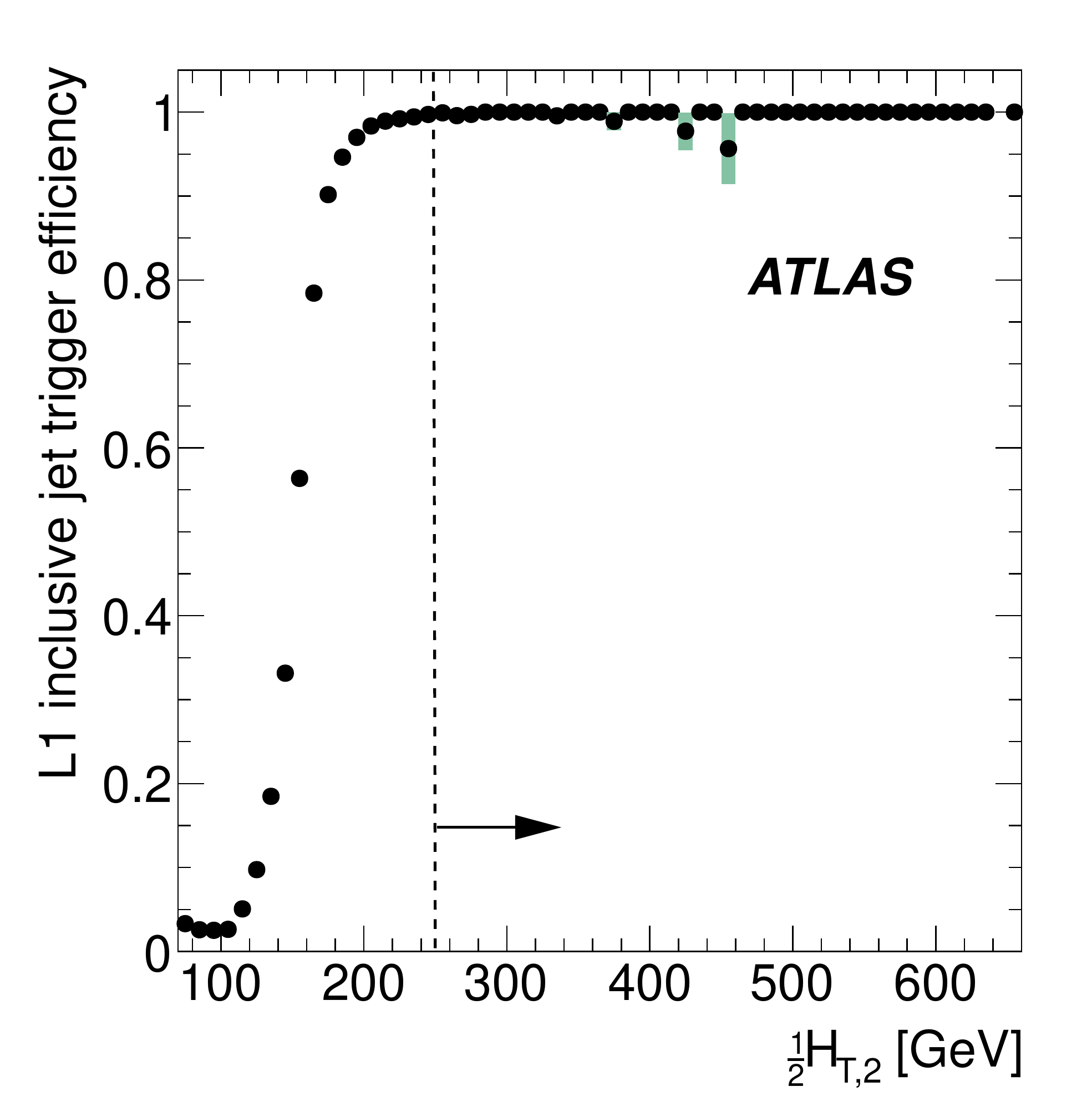}
  \caption{Inclusive jet trigger efficiency as a function of \HTtwo
           evaluated offline. The efficiency to select
           events with $\HTtwo > 250$~GeV using this trigger is greater
           than 99.8\%.
           }
  \label{fig:eventshapes:triggereff}
\end{figure}

A sample of events containing high-\pT\ jets is selected via a Level 1 inclusive single jet trigger with a nominal transverse energy threshold of 95~GeV at the EM scale. The offline selection requires two leading jets with a mean transverse momentum $\HTtwo > 250$~GeV and that the rapidity of each leading jet satisfy $|y|<1.0$. Subleading jets yield non-zero values of each event shape variable and must have $\pt>30$~GeV and be within $|y|<1.5$ in order to be used in the calculations. This choice of event selection is partially driven by the trigger threshold which is fully efficient only at an offline jet transverse momentum of $\pT>250$~GeV. High momentum jets are also less susceptible to the impact of multiple proton-proton interactions (pile-up) and have a smaller jet energy scale relative uncertainty. The use of \HTtwo has been shown to be significantly more stable against higher-order corrections to the jet cross section~\cite{Rubin2010}. The inclusive single jet trigger efficiency is evaluated with respect to the offline \HTtwo selection, as shown in Fig.~\ref{fig:eventshapes:triggereff}, and is on average greater than 99.8\%, thereby removing the need for trigger efficiency corrections. This determination is made \textit{in situ} using a trigger selection with a threshold of 30~GeV at the EM scale.

The presence of pile-up in these data has the potential to impact both event selection and jet reconstruction. Experimentally, reconstruction of primary vertices using tracks measured in the ID provides a measure of the multiplicity of such additional interactions on an event-by-event basis. The 2010 data contain an average number of primary vertices, \Npv, of approximately $\Navg=3$, with a tail extending to $\Npv\geq10$. The vertex with the highest total squared track momentum, $\sum p_{T, {\rm track}}^{2}$, is assumed to be the vertex at which the hard scattering that triggered the event occurred. 

Two primary effects are expected from pile-up: augmentation of the jet energy scale for jets produced in the hard scattering and pile-up jets produced directly by the additional \pp collisions within the same bunch crossing. The consequence of the former is typically an offset to the energy scale which is corrected as described below. The presence and impact of pile-up jets on the event shape measurements is discussed in more detail in Section~\ref{results:systematics}.

\subsection{Jet Reconstruction and Calibration}

Jets reconstructed with the \antikt\ algorithm~\cite{Cacciari:2008gp, Cacciari200657} are used for the event shape measurements presented here. This algorithm yields regular, approximately circular jets whose boundaries are well described by the nominal jet radius. A jet radius of $R=0.6$ is used here, as in many Standard Model jet-physics measurements in ATLAS, compared to the smaller radius $R=0.4$ jets used for a variety of new physics searches and top-quark measurements. This choice is made in order to minimize jet-by-jet corrections due to higher-order emissions and to maximize the reconstruction efficiency since the additional small uncertainties that accompany the larger radius do not significantly impact the measurement. Anti-$k_{t}$ jets have been shown to be less susceptible than other jet algorithms to systematic effects such as pile-up and close-by jet activity. The inputs to the jet algorithm are topological energy clusters at the EM scale~\cite{Lampl:1099735}. Following an average offset correction derived \textit{in situ} to account for noise and contributions due to pile-up, an $\eta$- and $\pT$-dependent jet energy correction referred to as the EM+JES correction~\cite{ATLASJESpaper2010} is applied to all jets to compensate for energy loss in the calorimeters, detector geometry and other effects. Jet quality criteria such as the timing of calorimeter cell signals, the EM energy fraction, and pulse shape information are used to select only those events with well measured jets. These criteria are designed to remove events that are likely to have contamination due to beam-related backgrounds, cosmic rays, or detector defects. In order to further reduce non-collision backgrounds, each event must contain at least one primary vertex consisting of at least five tracks with transverse momenta $\pttrk>150$~MeV. The MC simulation is reweighted in order to match the primary vertex multiplicity observed in the data. 

\subsection{Monte Carlo Simulation}

Dijet and multi-jet events are generated using two approaches. The first uses direct perturbative calculation of the tree-level matrix elements in powers of the strong coupling constant, \alphas. The matrix elements are evaluated at LO in \alphas\ for each relevant partonic subprocess. This is a so-called ``multi-leg'' method. The second approach implements a sampling of the phase space available for gluon emission with some suitable approximations. The latter uses LO perturbative calculations of matrix elements for \TwoToTwo processes and relies on the parton shower implementation to produce the equivalent of multi-parton final states.

The multi-leg technique is used by \Alpgen~\cite{alpgen}. In the analysis presented here, \Alpgen~2.13 is used with up to six final-state partons. \Alpgen is interfaced to both \Herwig~6.510~\cite{Corcella2001} to provide the parton shower and hadronization model, and to \Jimmy~4.31~\cite{jimmy} for the underlying event model. The CTEQ6L1 LO~\cite{PDF-CTEQ} parton distribution function (PDF) with LO \alphas\ is used for \Alpgen. 

The parton shower simulation programs \Pythia~\cite{Sjostrand2001} and \herwigpp~\cite{Bahr2008} both implement the second approach for QCD jet production and rely on the parton shower to generate multi-jet final states. \Pythia~6.423 with the Perugia 2010 tune~\cite{Perugia2010} and \Herwigpp~2.4.2 are used to compare to the data; these provide shower models that are $p_T$ ordered and angular ordered, respectively. LO PDFs are taken from the MRST2007 LO*~\cite{Martin:2009iq,Sherstnev:2007nd} PDF for \Herwigpp, and from the CTEQ6L1 LO~\cite{PDF-CTEQ} PDF in \Pythia. 

The MC programs used for comparison to the measurements of event shapes are chosen in part for their ability to describe other ATLAS jet-based measurements. The multi-jet cross section measurements~\cite{Collaboration:2011tq}, which constitute the same final states as those probed in this analysis, exhibit very good agreement with the predictions from \Alpgen. \Herwigpp not only exhibits good agreement with individual jet shape measurements~\cite{Collaboration:2011kq} but is also tuned to yield good agreement with event shape measurements from the Large Electron Positron (LEP) collider experiments. Finally, the Perugia 2010 tune of \Pythia also shows good agreement with the ATLAS jet shape measurements and has been tuned using the theoretical input from higher-order calculations of event shapes presented in Ref.~\cite{Banfi2010}. These three MC programs thus provide well motivated predictions for the final state observables measured via event shapes.

Events generated by these MC programs are passed through a full simulation~\cite{simulation} of the ATLAS detector and trigger based on \Geant~\cite{geant4} and processed in the same way as collision data. The Quark-Gluon String Precompound (QGSP) model~\cite{QGS} is used for high energy inelastic scattering of hadrons by nuclei, and the Bertini cascade model~\cite{Bertini} is used to describe the interactions of hadrons with the nuclear medium. Alternative \Geant physics lists that specify particle and process definitions, using a combination of the FRITIOF~\cite{Fritiof} and Bertini models and QGSP without Bertini, are used as part of the studies to understand the uncertainties on the jet energy scale.

%% file: sec-results.tex
The event shape measurements using jets presented here are corrected to particle-level, after accounting for detector efficiencies and instrumental effects. Particle-level jets are constructed from all final state particles from the MC simulation with lifetimes longer than 10~ps. Direct comparisons can thus be made between the results presented here and MC generator data after parton shower and hadronization. The dependence of these observables on \HTtwo is also evaluated. This allows both the investigation of the results and a clearer picture of regions of phase space in the measurement where MC predictions may not fully describe the data.

\subsection{Accounting for Detector Effects}
\label{results:unfolding}

In order to compare the predictions of MC event generators with the measurements, several effects must be accounted for.  Efficiency loss due to detector coverage and resolution, small detector biases, and other effects, may all affect the measured value of an event shape variable. In order to account for such effects, the MC and detector simulation are used to estimate their impact. MC events after full detector simulation are used to derive bin-by-bin corrections that are applied to the detector-level measurements of each event shape variable to obtain the unfolded, particle-level result to which the MC simulated events after parton shower may be compared. These corrections are typically smaller than 10\%. The bin sizes were chosen to be approximately commensurate with resolution, with individual bin purities significantly greater than 50\%.

The primary MC generator used for evaluating the corrections is \Alpgen, since the detector-level distributions are well described and \Alpgen models the multi-jet cross-section  well~\cite{Collaboration:2011tq}. As a cross-check of the method, the corrections evaluated with \Alpgen are compared to those obtained from \Herwigpp and \Pythia. For \lny, \apla, and \sph, this component of the uncertainty is approximately 2\%--8\%, which is smaller than both that due to the overall JES systematic uncertainty discussed below and the finite sample size with which the correction factors are determined. However, for the thrust event shape variables, in particular for \thrmin, and \sphtranv, the generator dependence of these corrections is approximately 10\% for the majority of the range of those measurements. 

\subsection{Systematic Uncertainties}
\label{results:systematics}

Multiple effects are present in the measurement of event shapes due to the inclusive nature of these observables. These include the uncertainty due to the jet energy scale, the effects of multiple \pp interactions, the finite resolution, and the fiducial range of the detector. All of these effects are evaluated and accounted for in the measurement. The dominant uncertainties are the jet energy scale and generator dependence of the corrections in regions of high statistical precision.

The uncertainty on the JES established by the jet calibration procedure~\cite{ATLASJESpaper2010} influences the final event shape measurement via both the thresholds used to select events and the momenta used to calculate the event shape observables. This uncertainty is primarily established by the measurement of the single hadron response using test beam data, but is also verified \textit{in situ} during 2010. For jets used in these measurements the typical JES uncertainties are 2.3\%--3.0\%. The impact of this source of systematic uncertainty is reduced by the explicit use of ratios of jet momenta for several observables, although the jet yield can still vary for a given event due to these selections. Variations of the individual jet momenta are performed within the systematic uncertainties of the JES measurement. For nearly all measured event shape variables, the overall JES uncertainty has the largest impact apart from statistical precision. Most observables have an approximate 5\% uncertainty due to the JES, with \apla and \thrmaj being impacted by up to 10\% in the steeply falling tails of the distributions.

Additional jets present in the event due to pile-up may also alter the observed event shape. This may be of particular importance for those measurements that are explicitly dependent upon the jet multiplicity, such as those computed from the event transverse momentum tensor. The impact of pile-up on the jet energy is accounted for by the energy scale corrections and uncertainty discussed above, where the pile-up contribution becomes negligible for $\pT>250$~GeV. 

A crucial tool in the identification of jets from pile-up is the jet-vertex fraction, or \JVF~\cite{ATLASWJetspaper2010}. This discriminant estimates the contribution of pile-up to a single jet by measuring the fraction of charged particle momentum in the jet that originates in the hard scatter. The rate of additional jets from pile-up in events with between five and eight primary vertices exhibits an increase of a factor of two compared to events with two primary vertices.  Because the overall fraction of events with greater than five reconstructed primary vertices is approximately two percent, and these jets tend to have a much softer \pT\ spectrum, the impact due to additional jets is significantly smaller than other systematic and statistical uncertainties for all measurements. 

To further establish the systematic uncertainty incurred by pile-up, comparisons are made between the observed detector-level distributions in events with and without additional reconstructed vertices. A slight variation of a few percent at low \lny is observed as well as a relative 10\% increase in the fraction of events at higher \thrmaj. Similar observations are made by evaluating the impact of the JES uncertainty. Furthermore, each event shape is measured as a function of the \JVF of the third jet in the event to directly test the effect of pile-up on the final state observables. Variations of less than 3\% are observed when requiring that jets contain a high fraction of associated track momentum originating in the identified hard-scatter vertex. This effect is taken into account in the systematic uncertainty in the final result.

\subsection{Event Shape Distributions}
\label{results:hadronlevel}

\begin{figure*}
  \begin{center}
    
      \subfigure[]{
	    \includegraphics[width=0.80\columnwidth]{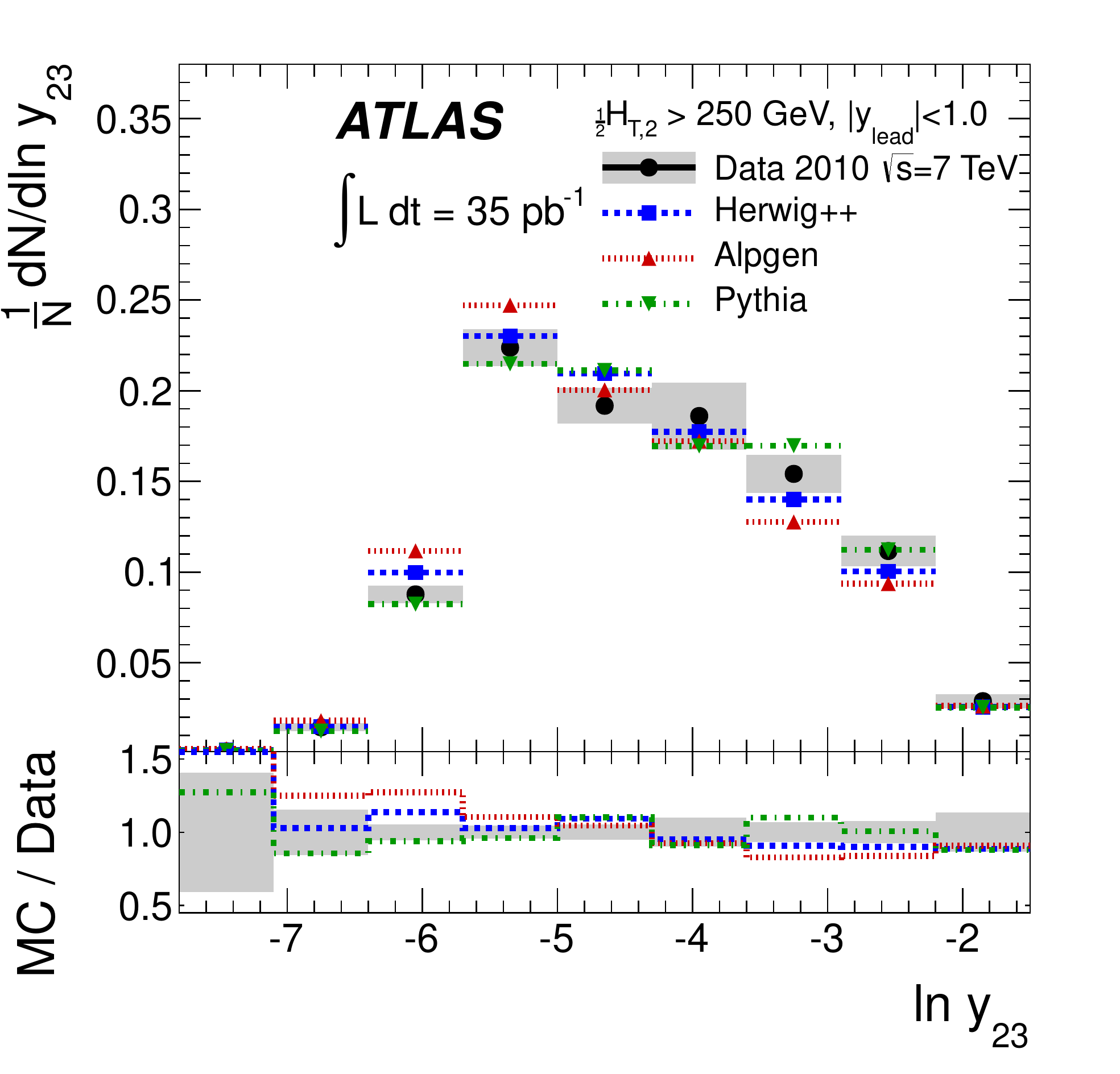}
	    \label{fig:eventshapes:unfolded:lny23}}
      \subfigure[]{
	    \includegraphics[width=0.80\columnwidth]{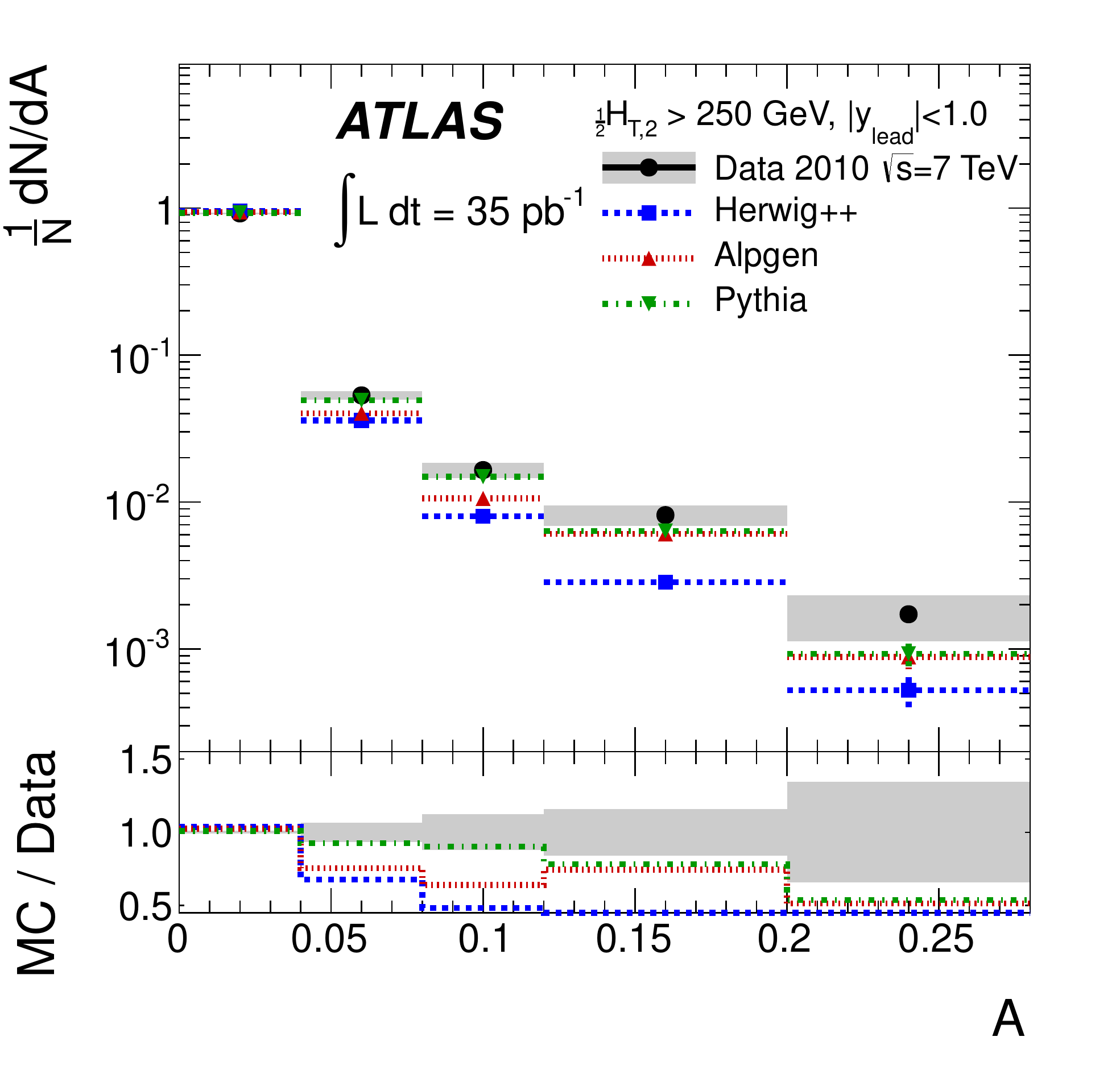}
	    \label{fig:eventshapes:unfolded:apl}}
    
	  \subfigure[]{
	    \includegraphics[width=0.80\columnwidth]{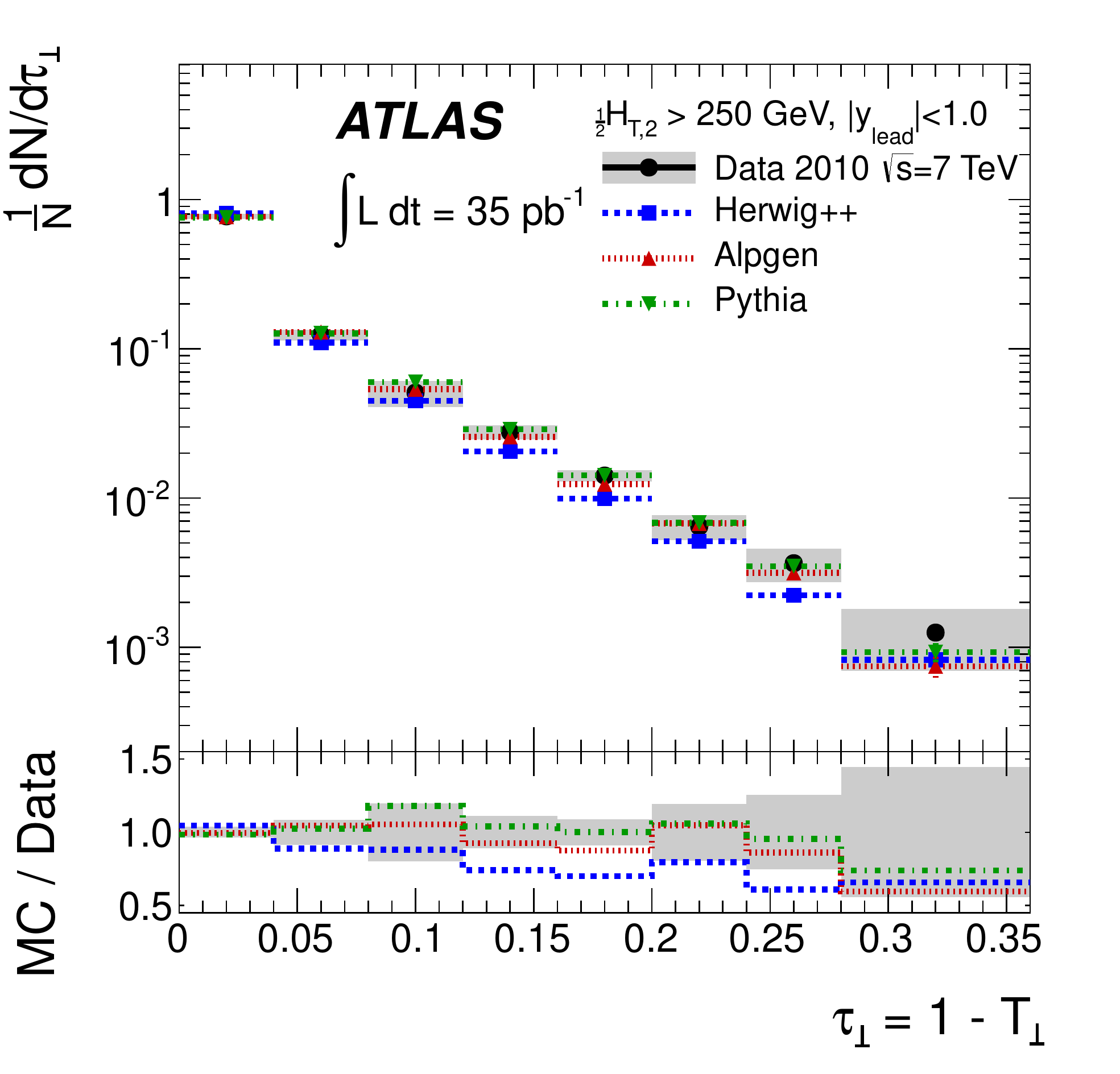}
	    \label{fig:eventshapes:unfolded:thrustMajor}}
	  \subfigure[]{
	    \includegraphics[width=0.80\columnwidth]{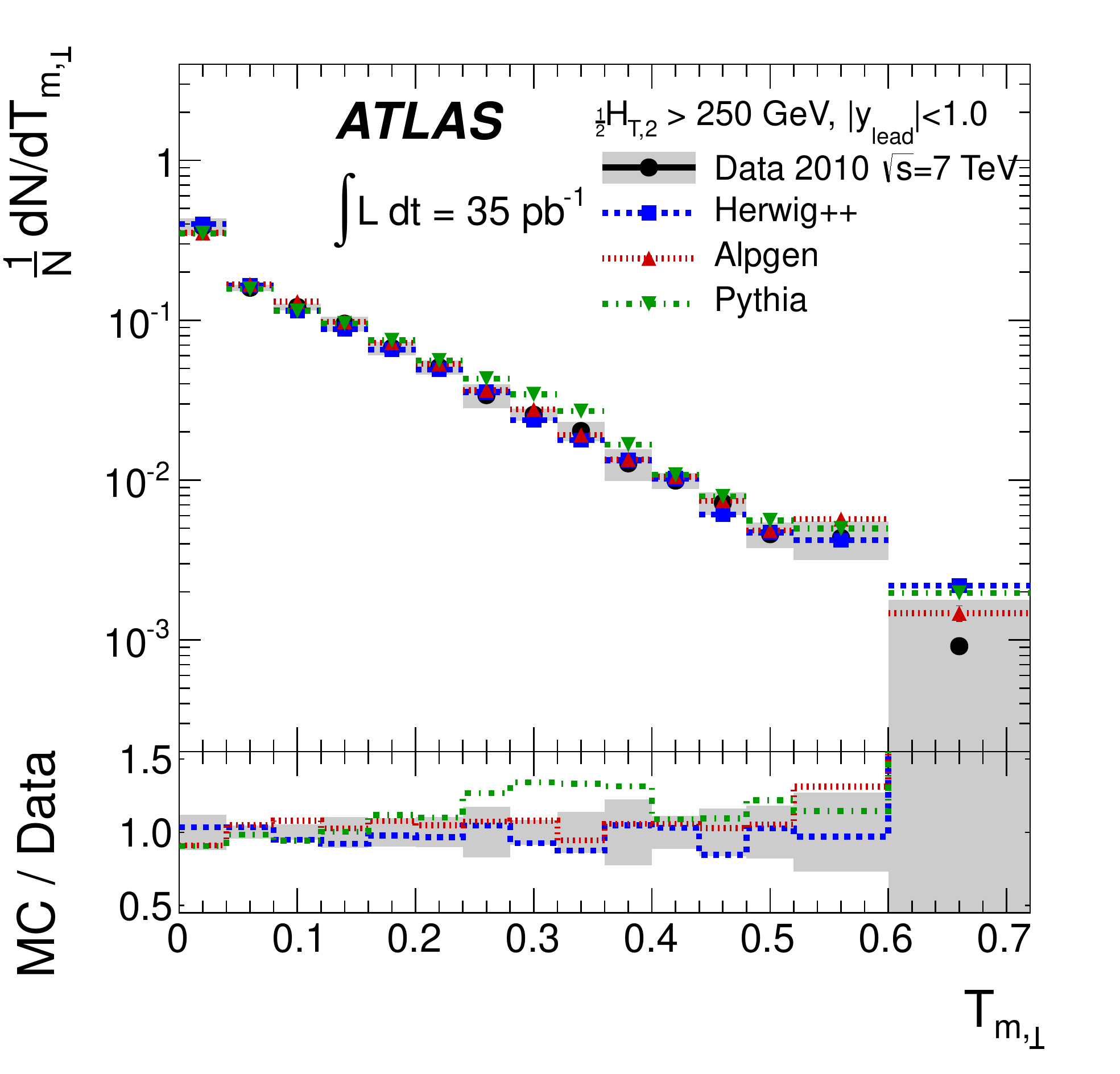}
	    \label{fig:eventshapes:unfolded:thrustMinor}}

	  \subfigure[]{
	    \includegraphics[width=0.80\columnwidth]{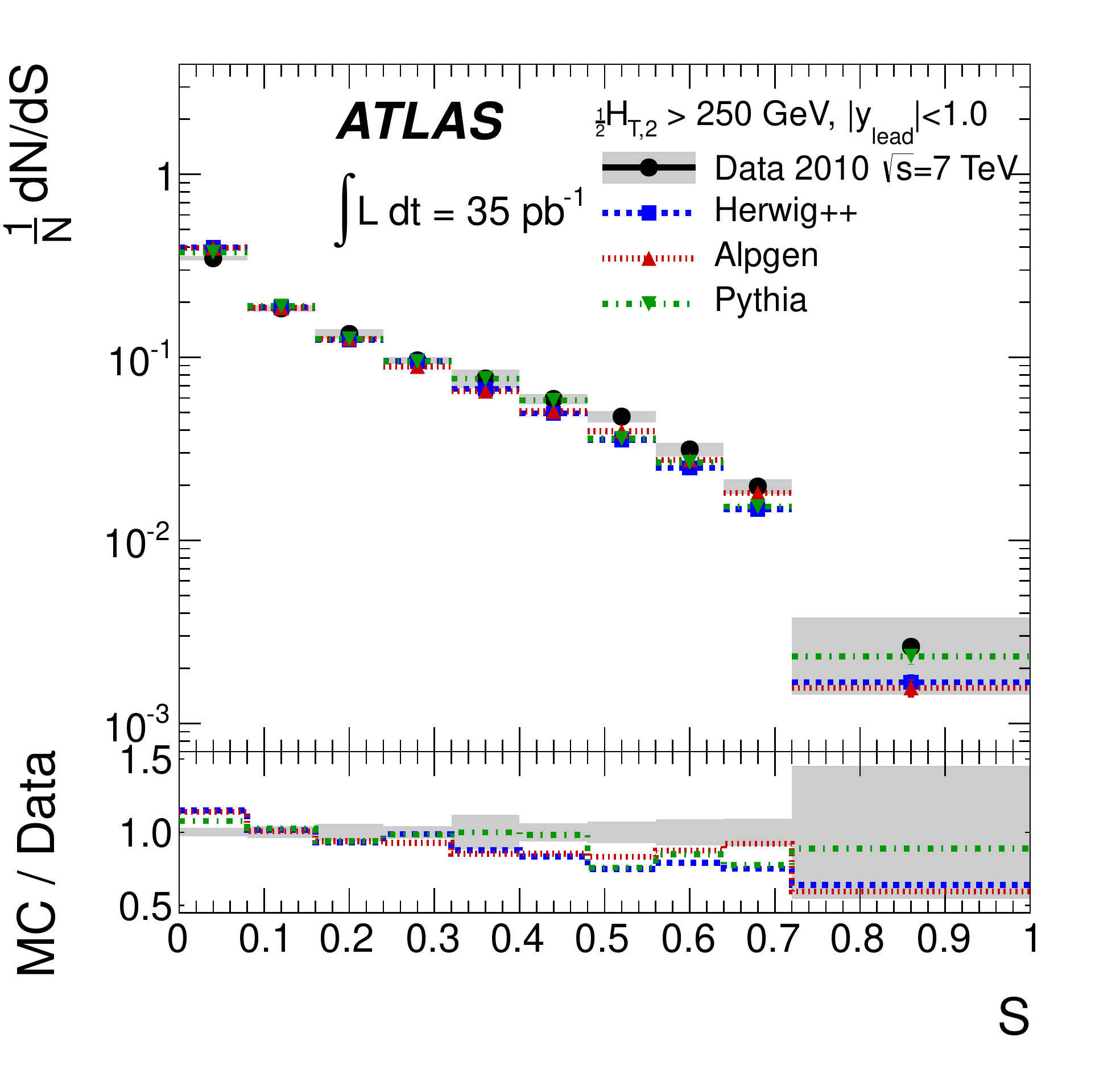}
	    \label{fig:eventshapes:unfolded:sph}}
	  \subfigure[]{
	    \includegraphics[width=0.80\columnwidth]{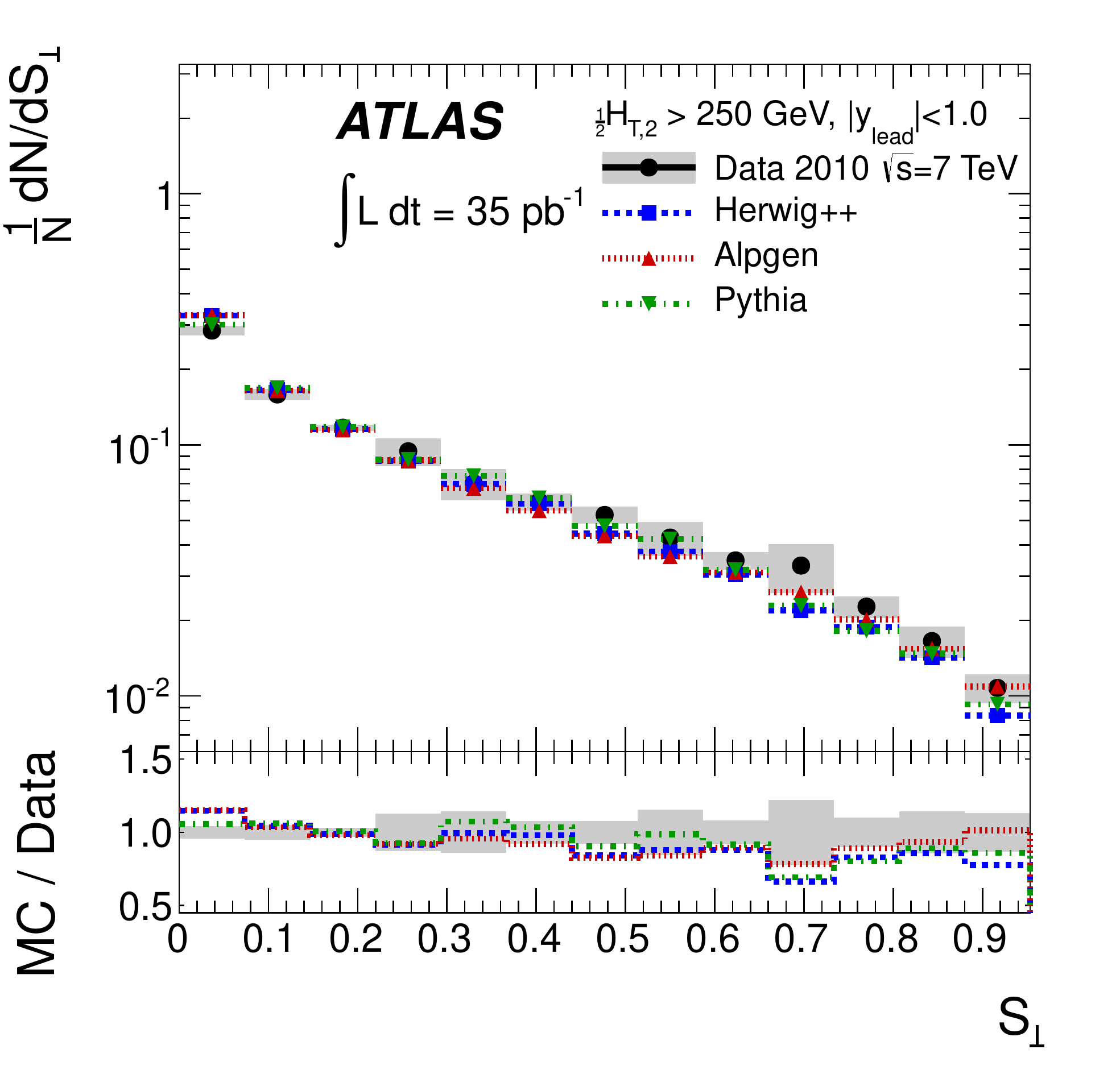}
	    \label{fig:eventshapes:unfolded:sphtranv}}
	    
  \end{center}
  
  \caption{Unfolded hadron-level distributions of the
           \subref{fig:eventshapes:unfolded:lny23} third-jet resolution parameter, \lny,
           \subref{fig:eventshapes:unfolded:apl} aplanarity, \apla,
           \subref{fig:eventshapes:unfolded:thrustMajor} transverse thrust, \thrmaj,
           \subref{fig:eventshapes:unfolded:thrustMinor} minor component of the 
           transverse thrust, \thrmin,
           \subref{fig:eventshapes:unfolded:sph} sphericity, \sph, and
           \subref{fig:eventshapes:unfolded:sphtranv} transverse sphericity, \sphtranv.
           The uncertainty shown for the data includes statistical and systematic uncertainties.}
  
  \label{fig:eventshapes:unfolded}
\end{figure*}

The normalized distributions of the third-jet resolution parameter and aplanarity are shown in Figs.~\ref{fig:eventshapes:unfolded:lny23} and~\subref{fig:eventshapes:unfolded:apl}. In the case of \ytwoth, where the primary sensitivity is to the description of the momentum of the third jet, \Pythia provides the most accurate description of the data, whereas \Herwigpp exhibits slightly better agreement than \Alpgen. Although \Alpgen provides exact tree-level matrix element calculations for up to six jets, it overestimates the fractions of events in the range $\lny<-5$. This is qualitatively expected because \Alpgen's more precise calculation of the high jet multiplicity states is primarily concerned with jets near the hard scale of the event. In this region, the impact of the leading-logarithm resummation in \Pythia and \Herwigpp plays a significant role. As a result, \Pythia and \Herwigpp both describe the data more accurately than \Alpgen for this event shape variable, particularly at small values of \lny. 

Aplanarity measures the sum of the transverse momenta out of the event plane defined primarily by the two hardest jets. The deviation of the MC prediction from the data is significant for \Herwigpp, with some small differences observed with respect to \Alpgen as well. The measurements consistently support more highly aplanar events than predicted by \Herwigpp, with the majority of the distribution observed to be significantly different from the MC prediction. The agreement with \Pythia is good across the full distribution. These results suggest that the event shape is more accurately described by the exact multi-jet prediction provided by the multi-leg matrix element generator (such as \Alpgen) and the model provided by \Pythia. 

The statement that the topological distribution is better described by \Alpgen and \Pythia than by \Herwigpp is supported by the measurement of the transverse thrust, \thrmaj. Fig.~\ref{fig:eventshapes:unfolded:thrustMajor} exhibits the same behavior as observed in the aplanarity: \Herwigpp predicts fewer than observed highly isotropic events at large \thrmaj. Throughout the distribution, \Alpgen and \Pythia both predict the measured thrust well. The minor component of the thrust, \thrmin, or the out-of-plane thrust magnitude, shown in Fig.~\ref{fig:eventshapes:unfolded:thrustMinor}, does not exhibit as large a difference as observed in the aplanarity. A slight overestimation by \Pythia is observed for intermediate values of the thrust minor component, $0.25<\thrmin<0.40$.

Lastly, the sphericity and transverse sphericity distributions shown in Figs.~\ref{fig:eventshapes:unfolded:sph} and~\subref{fig:eventshapes:unfolded:sphtranv} exhibit differences between all three generators and the data. The construction of the transverse sphericity as a ratio of eigenvalues of the momentum tensor of the event leads to a slightly improved description. In both cases, \Pythia provides the best description of the data, and the data are better described by \Alpgen than \Herwigpp.  In particular, \Herwigpp underestimates the number of highly spherical events in the range $0.40 < \sph < 0.72$.

\subsection{Dependence on \HTtwo}
\label{results:kinematics}

The measurement of the distribution of event shape observables allows for a detailed comparison with MC predictions for a large range of kinematic phase space defined by \HTtwo in multi-jet events. It is informative to evaluate the explicit dependence of these shapes on the kinematic properties of the event in order to determine potential differences in the modeling of the data for different jet momentum ranges. The evolution of each event shape variable with \HTtwo exhibits a similar trend as that expected by the running of \alphas which leads to a reduction in extra gluon radiation and thus a reduction in the value of each event shape. 

\begin{figure*}
  \begin{center}
    
      \subfigure[]{
        \includegraphics[width=0.80\columnwidth]{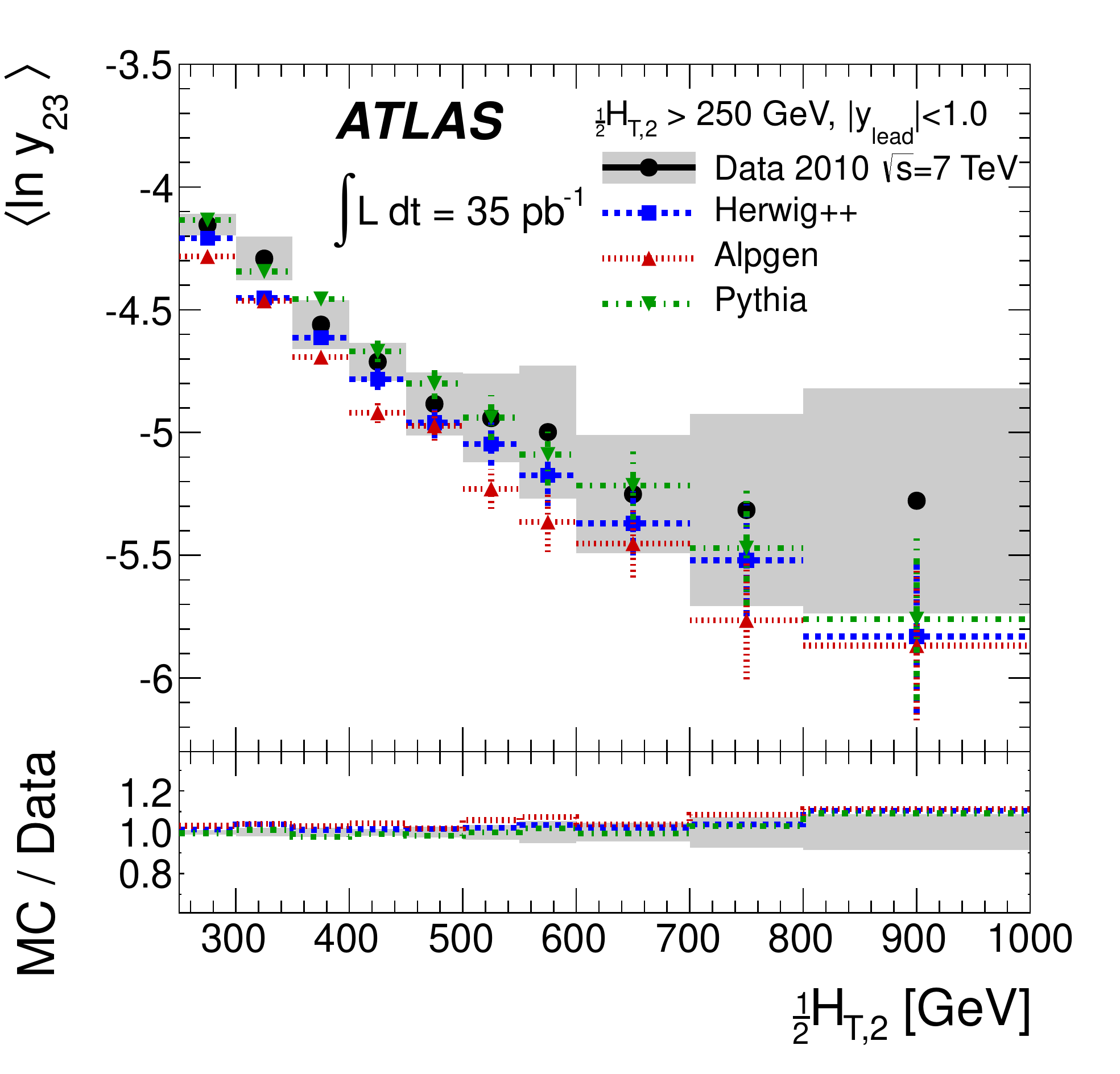}
        \label{fig:eventshapes:y23HT2:noPU}}
      \subfigure[]{
        \includegraphics[width=0.80\columnwidth]{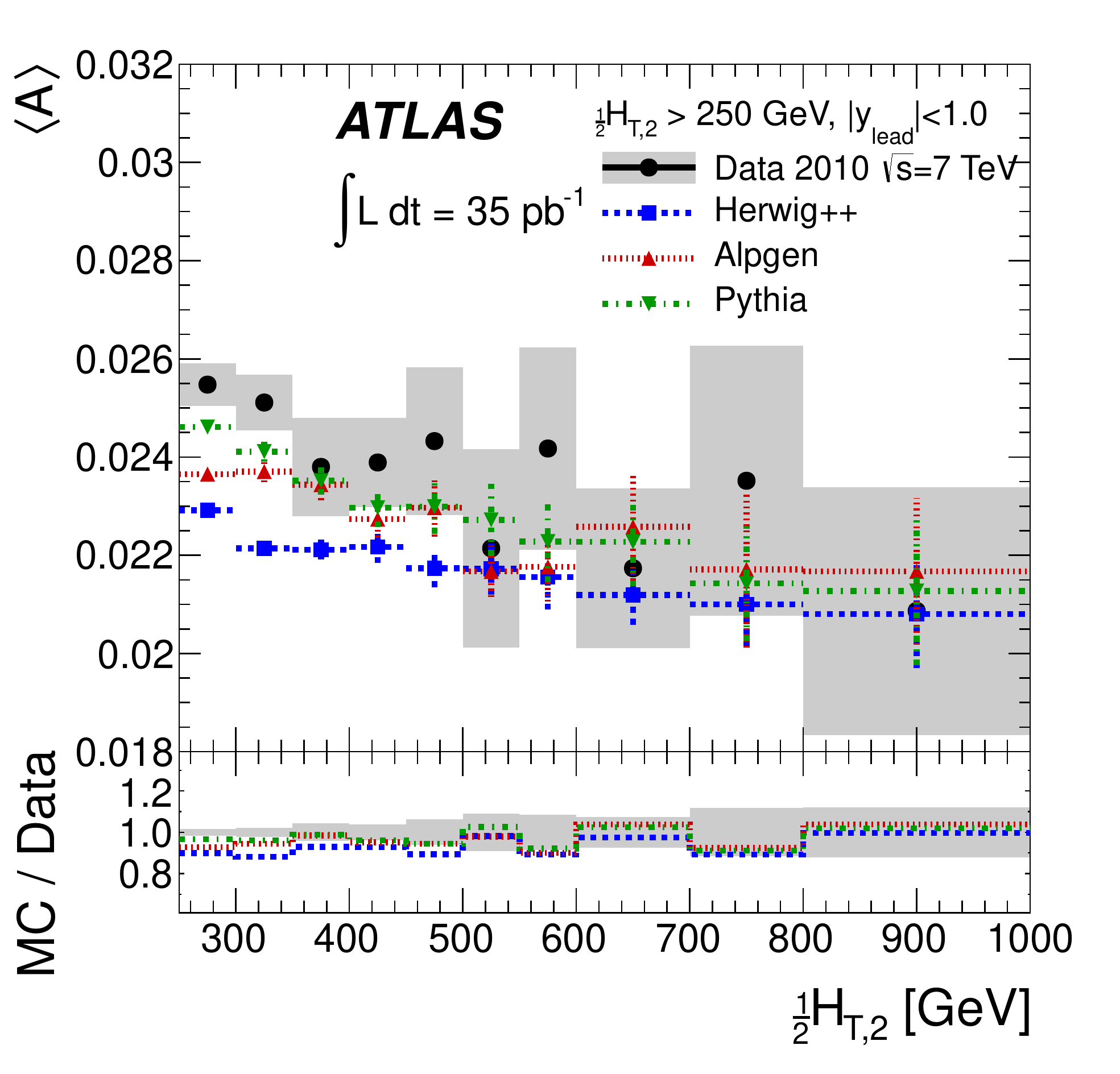}
        \label{fig:eventshapes:aplHT2:noPU}}
    
	  \subfigure[]{
	    \includegraphics[width=0.80\columnwidth]{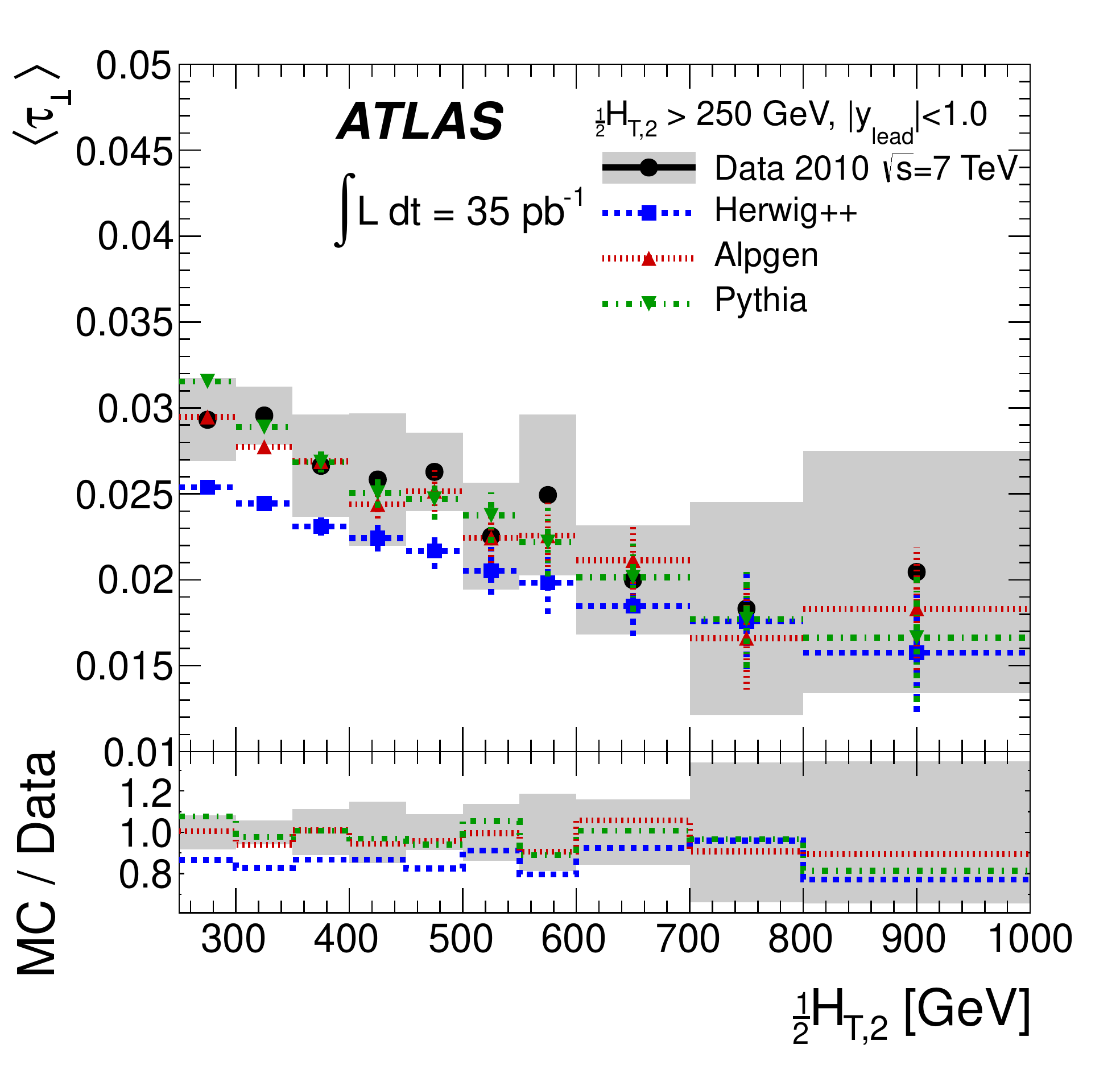}
	    \label{fig:eventshapes:thrmajHT2:noPU}}
	  \subfigure[]{
	    \includegraphics[width=0.80\columnwidth]{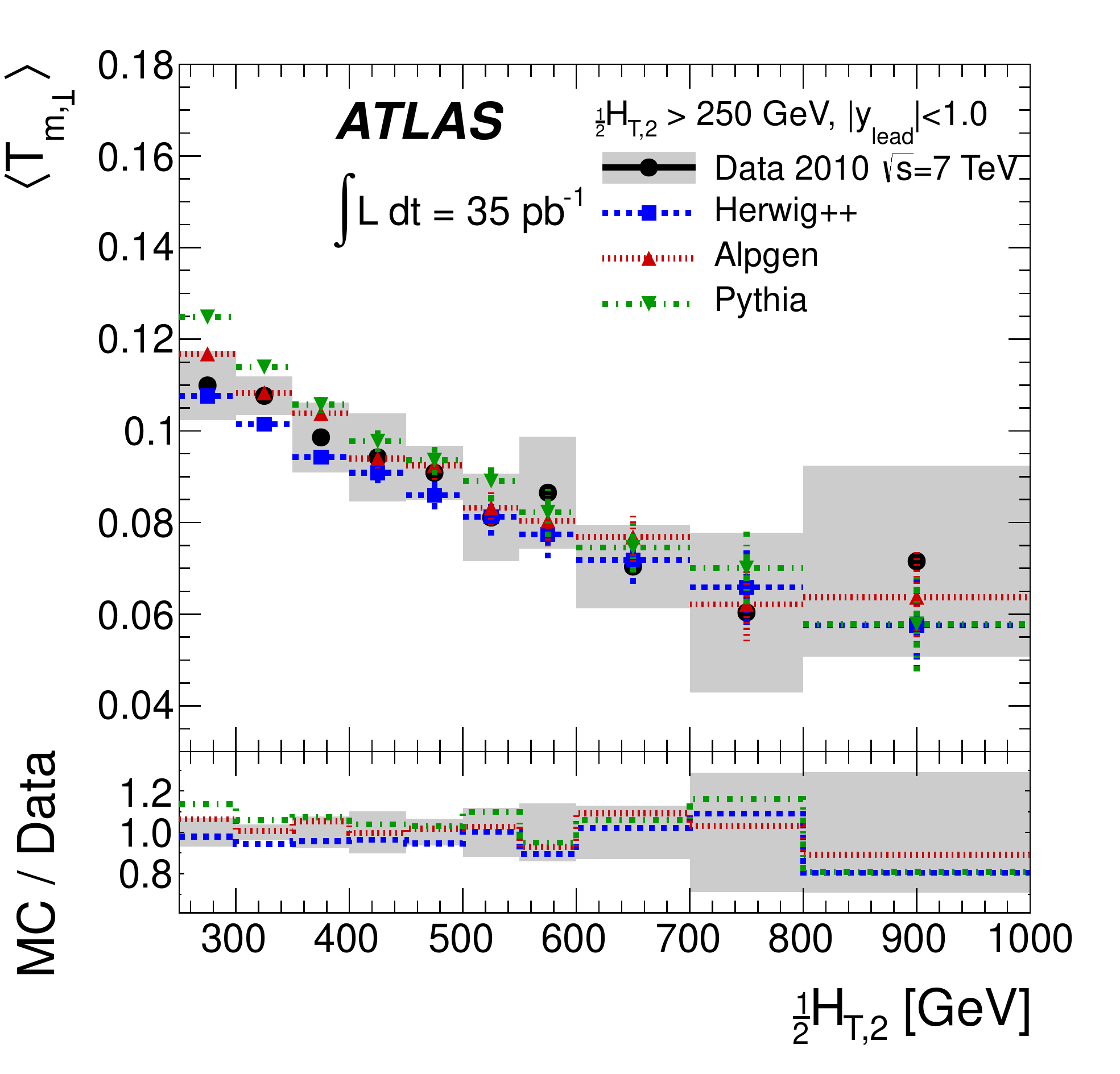}
	    \label{fig:eventshapes:thrminHT2:noPU}}

	  \subfigure[]{
	    \includegraphics[width=0.80\columnwidth]{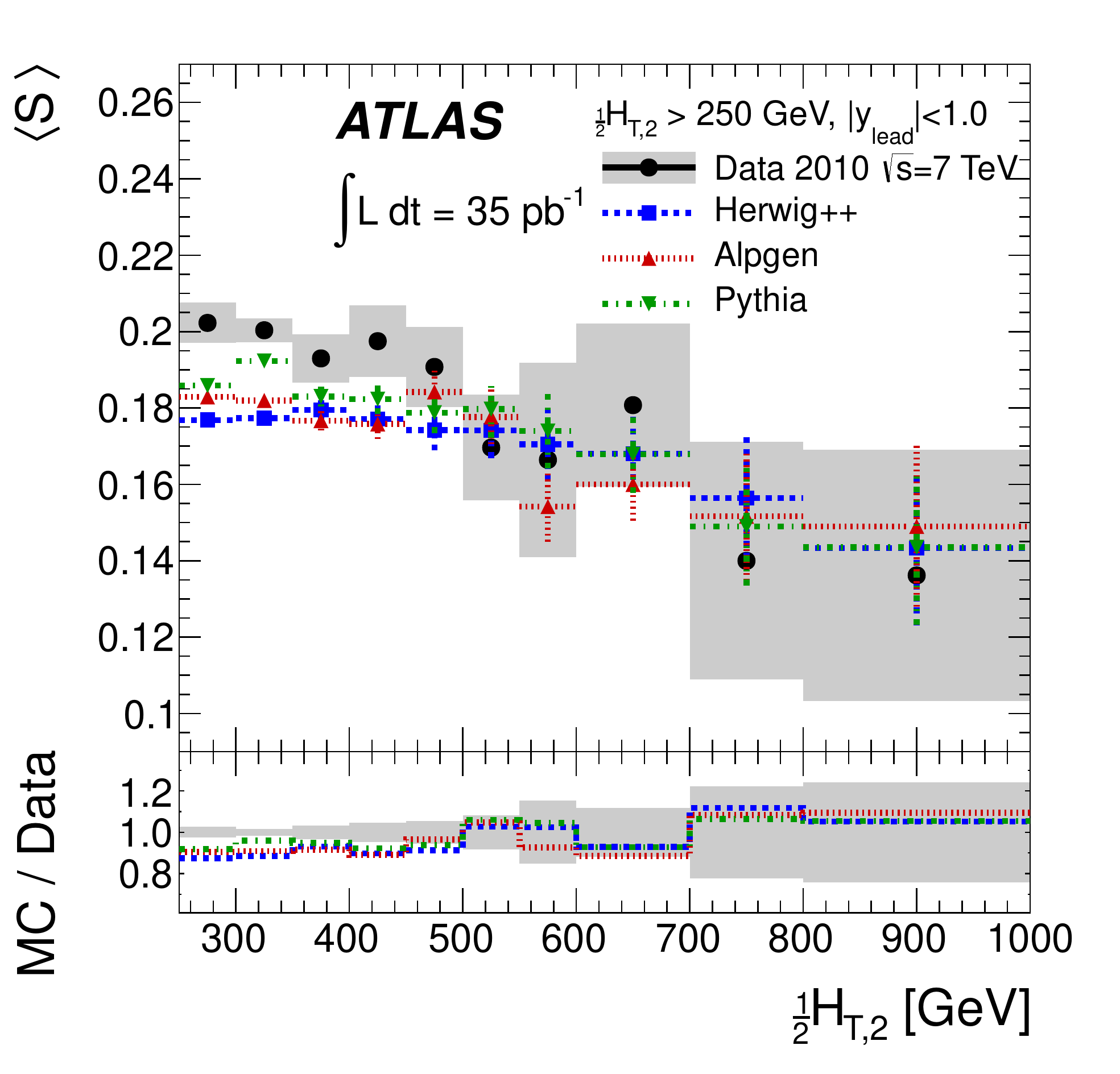}
	    \label{fig:eventshapes:sphHT2:noPU}}
	  \subfigure[]{
	    \includegraphics[width=0.80\columnwidth]{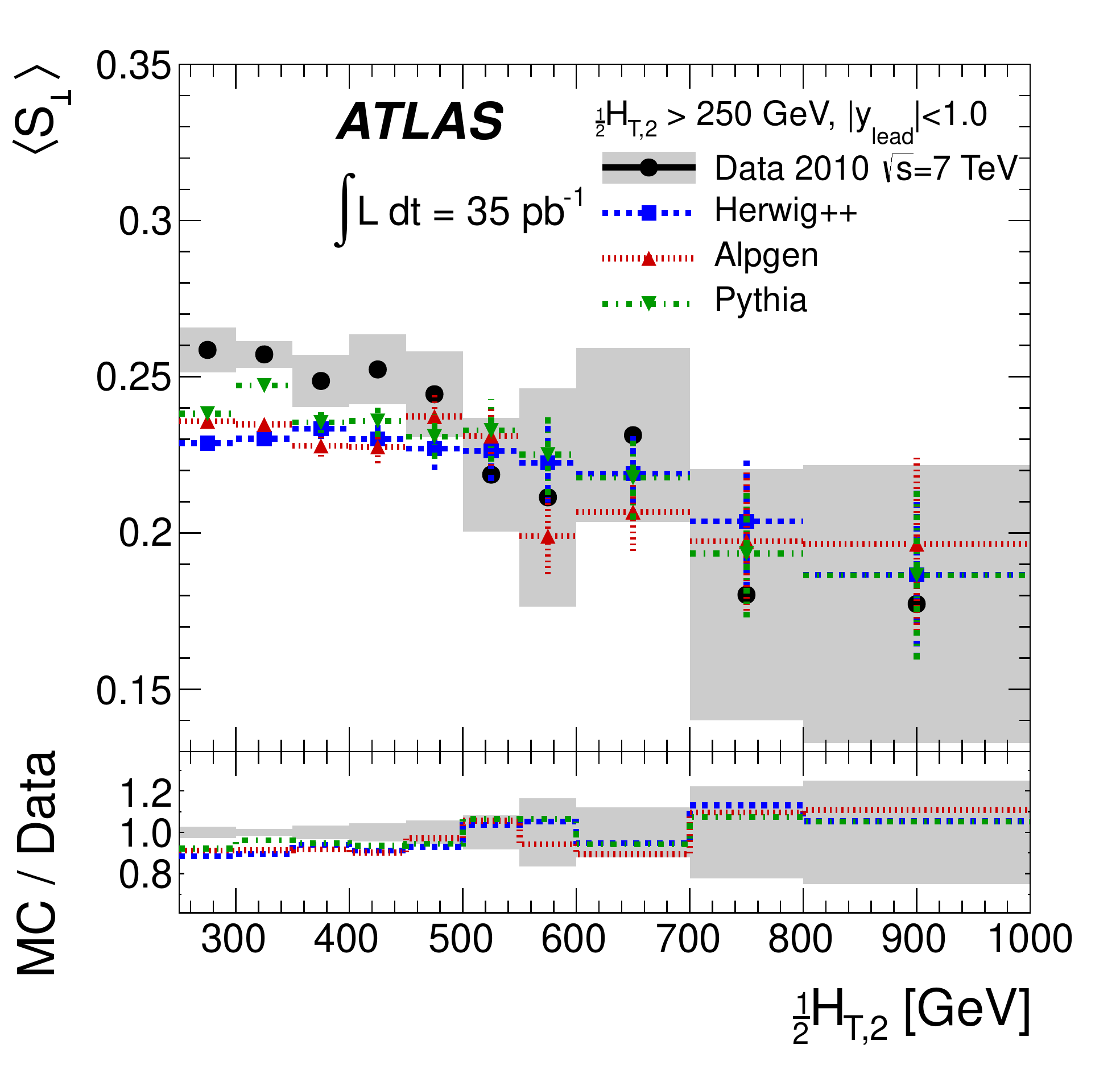}
	    \label{fig:eventshapes:sphTHT2:noPU}}

  \end{center}
  
  \caption{Mean value of each event shape variable as a function of \HTtwo.
           Comparisons are made between the 
           MC generators \Herwigpp, \Alpgen and \Pythia.}
  \label{fig:eventshapes:HT2dependence}
\end{figure*}

Figure~\ref{fig:eventshapes:HT2dependence} depicts the dependence of the mean of each event shape variable on \HTtwo as it approaches 1~TeV. In all cases, a general trend is observed in which the mean decreases as \HTtwo increases. This can be understood in terms of how the dynamics of the \TwoToTwo process evolves with energy. As the energy in the leading jets increases, the di-jet structure dominates because of kinematics and because \alphas\ decreases as \HTtwo increases, causing the dominant \NLO corrections which generate higher relative momentum gluon emission to decrease as well. 

The variation as a function of \HTtwo is the largest for \meanlny, which indicates a change of \ytwoth of nearly a factor of five between $\HTtwo=300$~GeV and 800~GeV. Nonetheless, the agreement between the MC prediction and the observed event shape variable dependence is good for all generators. The observation made above that too few highly aplanar events are present in the predictions from \Herwigpp is again observed in the evolution of \meanapla with the momentum scale of the event. The agreement among the mean values measured and the MC predictions improves for $\HTtwo>500$~GeV, although the systematic uncertainties are larger and the statistical power of the measurement is reduced. Similarly, the evolution of $\langle\thrmaj\rangle$ with \HTtwo in Fig.~\ref{fig:eventshapes:thrmajHT2:noPU} is underestimated by \Herwigpp. \Alpgen and \Pythia consistently predict the mean value of \thrmaj and its evolution with \HTtwo more accurately, whereas all of the generators describe $\langle\thrmin\rangle$ vs. \HTtwo well. Finally, the sphericity and the transverse sphericity (Figs.~\ref{fig:eventshapes:sphHT2:noPU} and~\subref{fig:eventshapes:sphTHT2:noPU}) are both measured to be approximately 10\% larger than predicted by the three MC programs at low \HTtwo, whereas the agreement again improves for higher values. This difference is driven by the underestimate of highly spherical events observed in Figs.~\ref{fig:eventshapes:unfolded:sph} and~\subref{fig:eventshapes:unfolded:sphtranv} which decreases as the average sphericity decreases as a function of \HTtwo. 

%% file: sec-conclusions.tex
Six event shape observables 
are measured with jets in proton-proton collisions at \sqsseven with the ATLAS detector with a data sample of 35\invpb. 

Measurements are performed up to an event $H_{T,2}$ of 2~TeV and are compared with different Monte Carlo event generators. Overall shape comparisons are made with these MC programs, as well as the kinematic evolution of the mean value of each event shape variable with \HTtwo. Reasonable agreement is observed in most kinematic and topological regions. The measurements suggest that the modeling of the data by \PythiaP and \Alpgen are more accurate than that by \Herwigpp, in particular for the aplanarity, \apla, and transverse thrust, \thrmaj. The good description provided by \Alpgen(+\Herwig/\Jimmy) of the multi-jet cross-section~\cite{Collaboration:2011tq} is reflected as well in these measurements, although the description provided by \Pythia tends to model the data more accurately. \Pythia predicts a slightly higher mean \thrmin at low \HTtwo than observed in the data whereas \Herwigpp predicts a slightly lower mean thrust. The systematic uncertainties in the measurement of \sph and \sphtranv are found to be relatively small. The observation that the measured mean value of each event shape variable decreases with \HTtwo is consistent with the trend expected from the running of \alphas and is generally well modeled by the MC simulations.

Comparisons of these results to previously published LHC measurements of hadronic event shapes~\cite{Collaboration2011g} indicate that the slight overestimate by \Pythia of events with thrust minor component, \thrmin, in the range $0.25<\thrmin<0.40$ is observed in each case. However, the good agreement observed in this study between data and both \Pythia and \Alpgen for the thrust, \thrmaj is not seen in Ref.~\cite{Collaboration2011g}. The different tunes of \Herwigpp and \Pythia, as well as the different underlying event and hadronization models interfaced to \Alpgen in the two measurements, may account for these differences. Furthermore, the systematic uncertainties associated with these measurements are in many cases similar to the differences between the generators.

ATLAS measurements of the jet shape~\cite{Collaboration:2011kq}, jet fragmentation function~\cite{ATLASJetFragmentation2010}, and multi-jet cross-section also provide additional insight into the results shown here. \Pythia and \Herwigpp 2.4.2 both provide a reasonable description of the fragmentation function of high momentum jets in the data, whereas only the former models the jet shape accurately. \Alpgen, on the other hand, yields a fairly accurate description of the three-jet to two-jet cross-section ratio~\cite{Collaboration:2011tq}, although it produces jet shapes that are significantly narrower than those measured here. 

These results, supported by other ATLAS measurements of the hadronic final state, reinforce the importance of the leading-logarithm resummation in parton shower MC event generators. They also demonstrate the ability of leading order MC to provide a reasonable description of multi-jet event shapes.

%% file: sec-acknowledgments.tex
We thank CERN for the very successful operation of the LHC, as well as the
support staff from our institutions without whom ATLAS could not be
operated efficiently.

We acknowledge the support of ANPCyT, Argentina; YerPhI, Armenia; ARC,
Australia; BMWF, Austria; ANAS, Azerbaijan; SSTC, Belarus; CNPq and FAPESP,
Brazil; NSERC, NRC and CFI, Canada; CERN; CONICYT, Chile; CAS, MOST and NSFC,
China; COLCIENCIAS, Colombia; MSMT CR, MPO CR and VSC CR, Czech Republic;
DNRF, DNSRC and Lundbeck Foundation, Denmark; EPLANET and ERC, European Union;
IN2P3-CNRS, CEA-DSM/IRFU, France; GNAS, Georgia; BMBF, DFG, HGF, MPG and AvH
Foundation, Germany; GSRT, Greece; ISF, MINERVA, GIF, DIP and Benoziyo Center,
Israel; INFN, Italy; MEXT and JSPS, Japan; CNRST, Morocco; FOM and NWO,
Netherlands; RCN, Norway; MNiSW, Poland; GRICES and FCT, Portugal; MERYS
(MECTS), Romania; MES of Russia and ROSATOM, Russian Federation; JINR; MSTD,
Serbia; MSSR, Slovakia; ARRS and MVZT, Slovenia; DST/NRF, South Africa;
MICINN, Spain; SRC and Wallenberg Foundation, Sweden; SER, SNSF and Cantons of
Bern and Geneva, Switzerland; NSC, Taiwan; TAEK, Turkey; STFC, the Royal
Society and Leverhulme Trust, United Kingdom; DOE and NSF, United States of
America.

The crucial computing support from all WLCG partners is acknowledged
gratefully, in particular from CERN and the ATLAS Tier-1 facilities at
TRIUMF (Canada), NDGF (Denmark, Norway, Sweden), CC-IN2P3 (France),
KIT/GridKA (Germany), INFN-CNAF (Italy), NL-T1 (Netherlands), PIC (Spain),
ASGC (Taiwan), RAL (UK) and BNL (USA) and in the Tier-2 facilities
worldwide.

%% file: atlas_authlist.tex
\begin{flushleft}
{\Large The ATLAS Collaboration}

\bigskip

G.~Aad$^{\rm 48}$,
B.~Abbott$^{\rm 111}$,
J.~Abdallah$^{\rm 11}$,
S.~Abdel~Khalek$^{\rm 115}$,
A.A.~Abdelalim$^{\rm 49}$,
O.~Abdinov$^{\rm 10}$,
B.~Abi$^{\rm 112}$,
M.~Abolins$^{\rm 88}$,
O.S.~AbouZeid$^{\rm 158}$,
H.~Abramowicz$^{\rm 153}$
% ...
\bigskip

% ...
$^{10}$ Institute of Physics, Azerbaijan Academy of Sciences, Baku, Azerbaijan\\
$^{11}$ Institut de F\'isica d'Altes Energies and Departament de F\'isica de la Universitat Aut\`onoma  de Barcelona and ICREA, Barcelona, Spain\\
% ...
$^{48}$ Fakult\"{a}t f\"{u}r Mathematik und Physik, Albert-Ludwigs-Universit\"{a}t, Freiburg i.Br., Germany\\
$^{49}$ Section de Physique, Universit\'e de Gen\`eve, Geneva, Switzerland\\
% ...
$^{88}$ Department of Physics and Astronomy, Michigan State University, East Lansing MI, United States of America\\
% ...
$^{111}$ Homer L. Dodge Department of Physics and Astronomy, University of Oklahoma, Norman OK, United States of America\\
$^{112}$ Department of Physics, Oklahoma State University, Stillwater OK, United States of America\\
% ...
$^{115}$ LAL, Universit\'e Paris-Sud and CNRS/IN2P3, Orsay, France\\
% ...
$^{153}$ Raymond and Beverly Sackler School of Physics and Astronomy, Tel Aviv University, Tel Aviv, Israel\\
% ...
$^{158}$ Department of Physics, University of Toronto, Toronto ON, Canada\\
% ...
\end{flushleft}


%% file: eventshapes-paper-epjc.bbl
\providecommand{\href}[2]{#2}\begingroup\raggedright\begin{thebibliography}{10}

\bibitem{OpalEventShapes2005}
{OPAL Collaboration}, {\em {Measurement of event shape distributions and
  moments in $e^+e^- \rightarrow$ hadrons at 91-209~GeV and a determination of
  $\alpha_{s}$}\/},  \href{http://dx.doi.org/10.1140/epjc/s2005-02120-6}{Eur.
  Phys. J. C {\bf 40} (2005)  287--316}.

\bibitem{H1EventShapes2006}
{H1 Collaboration}, {\em {Event shapes in deep inelastic scattering at
  HERA}\/},  \href{http://dx.doi.org/10.1140/epjc/s2006-02493-x}{Eur. Phys. J.
  C {\bf 46} (2006)  343}.

\bibitem{ZeusEventShapes2007}
{ZEUS Collaboration}, {\em {Event shapes in deep inelastic scattering at
  HERA}\/},  \href{http://dx.doi.org/10.1016/j.nuclphysb.2006.05.016}{Nucl.
  Phys. B {\bf 767} (2007)  1}.

\bibitem{JadeEventShapes2009}
{{JADE}} Collaboration, {\em {Determination of the Strong Coupling $\alpha_s$
  from hadronic Event Shapes and NNLO QCD predictions using JADE Data}\/},
  \href{http://dx.doi.org/10.1140/epjc/s10052-009-1149-1}{Eur. Phys. J. C {\bf
  64} (2009)  351}.

\bibitem{Kluth2001}
S.~Kluth et al., {\em {A measurement of the QCD colour factors using event
  shape distributions at $\sqrt{s}=14$ to 189 GeV}\/},
  \href{http://dx.doi.org/10.1007/s100520100742}{Eur. Phys. J. C {\bf 21}
  (2001)  199}.

\bibitem{Stewart2010}
I.~W. Stewart, F.~J. Tackmann, and W.~J. Waalewijn, {\em {N-Jettiness: An
  Inclusive Event Shape to Veto Jets}\/},
  \href{http://arxiv.org/abs/1004.2489}{{\tt arXiv:1004.2489 [hep-ph]}}.

\bibitem{Collaboration2011g}
{CMS Collaboration}, {\em {First measurement of hadronic event shapes in $pp$
  collisions at $\sqrt{s}=7$~TeV}\/},
  \href{http://dx.doi.org/10.1016/j.physletb.2011.03.060}{Phys. Lett. B {\bf
  699} (2011)  48}.

\bibitem{Webber:1998um}
B.~R. Webber, {\em {QCD power corrections from a simple model for the running
  coupling}\/},
\href{http://dx.doi.org/10.1088/1126-6708/1998/10/012}{JHEP {\bf 10} (1998)
  012}.

\bibitem{Anchordoqui2010}
L.~A. Anchordoqui et al., {\em {Searching for the Layered Structure of Space at
  the LHC}\/},  \href{http://dx.doi.org/10.1103/PhysRevD.83.114046}{Phys. Rev.
  D {\bf 83} (2011)  114046}.

\bibitem{Banfi2004}
A.~Banfi, G.~P. Salam, and G.~Zanderighi, {\em {Resummed event shapes at
  hadron-hadron colliders}\/},
  \href{http://dx.doi.org/10.1088/1126-6708/2004/08/062}{JHEP {\bf 08} (2004)
  062}.

\bibitem{Banfi2010}
A.~Banfi, G.~P. Salam, and G.~Zanderighi, {\em {Phenomenology of event shapes
  at hadron colliders}\/},
  \href{http://dx.doi.org/10.1007/JHEP06(2010)038}{JHEP {\bf 06} (2010)  038}.

\bibitem{Akrawy1990389}
{OPAL Collaboration}, {\em A study of jet production rates and a test of QCD on
  the Z0 resonance\/},
  \href{http://dx.doi.org/10.1016/0370-2693(90)91983-I}{Phys. Lett. B {\bf 235}
  (1990)  389}.

\bibitem{Bartel1982}
{JADE} Collaboration, {\em {Observation of four-jet structure in $e^{+}e^{-}$
  annihilation at $\sqrt{s}= 33$~GeV}\/},
  \href{http://dx.doi.org/10.1016/0370-2693(82)90383-5}{Phys. Lett. B {\bf 115}
  (1982)  338}.

\bibitem{Kluth2009}
S.~Kluth, {\em {$\alpha_s(M_Z)$ from Jade Event Shapes}\/},
  \href{http://arxiv.org/abs/0905.4891}{{\tt arXiv:0905.4891 [hep-ex]}}.

\bibitem{Bjorken1970}
J.~Bjorken and S.~Brodsky, {\em {Statistical Model for Electron-Positron
  Annihilation into Hadrons}\/},
  \href{http://dx.doi.org/10.1103/PhysRevD.1.1416}{Phys. Rev. D {\bf 1} (1970)
  1416}.

\bibitem{Barate19981}
{ALEPH} Collaboration, {\em {Studies of Quantum Chromodynamics with the ALEPH
  detector}\/},  \href{http://dx.doi.org/10.1016/S0370-1573(97)00045-8}{Physics
  Reports {\bf 294} (1998)  1}.

\bibitem{Brandt1964}
S.~Brandt, C.~Peyrou, R.~Sosnowski, and A.~Wroblewski, {\em {The principal axis
  of jets -- an attempt to analyse high-energy collisions as two-body
  processes}\/},  \href{http://dx.doi.org/10.1016/0031-9163(64)91176-X}{Physics
  Letters {\bf 12} (1964)  57}.

\bibitem{Bartel1986}
{JADE} Collaboration, {\em {Experimental studies on multijet production in
  $e^{+}e^{-}$ annihilation at PETRA energies}\/},
  \href{http://dx.doi.org/10.1007/BF01410449}{Z. Phys. C {\bf 33} (1986)  23}.

\bibitem{detPaper}
{ATLAS Collaboration}, {\em {The ATLAS Experiment at the CERN Large Hadron
  Collider}\/},
\href{http://dx.doi.org/10.1088/1748-0221/3/08/S08003}{JINST {\bf 3} (2008)
  S08003}.

\bibitem{PerfWithData2010}
{ATLAS Collaboration}, {\em {Performance of the ATLAS Detector using First
  Collision Data}\/},
\href{http://dx.doi.org/10.1007/JHEP09(2010)056}{JHEP {\bf 09} (2010)  056}.

\bibitem{InDetPerfPaper}
{ATLAS Collaboration}, {\em {The ATLAS Inner Detector commissioning and
  calibration}\/},
\href{http://dx.doi.org/10.1140/epjc/s10052-010-1366-7}{Eur. Phys. J. C {\bf
  70} (2010)  787}.

\bibitem{LArReadiness}
{ATLAS Collaboration}, {\em {Readiness of the {ATLAS} liquid argon calorimeter
  for {LHC} collisions}\/},
  \href{http://dx.doi.org/10.1140/epjc/s10052-010-1354-y}{Eur. Phys. J. C {\bf
  70} (2010)  723}.

\bibitem{TileReadiness}
{ATLAS Collaboration}, {\em {Readiness of the {ATLAS} Tile Calorimeter for
  {LHC} collisions}\/},
  \href{http://dx.doi.org/10.1140/epjc/s10052-010-1508-y}{Eur. Phys. J. C {\bf
  70} (2010)  1193}.

\bibitem{TriggerPerf2010}
{ATLAS Collaboration}, {\em {Performance of the ATLAS Trigger System in
  2010}\/},  \href{http://dx.doi.org/10.1140/epjc/s10052-011-1849-1}{Eur. Phys.
  J. C {\bf 72} (2012)  1849}.

\bibitem{ATLASJESpaper2010}
{ATLAS Collaboration}, {\em {Jet energy measurement with the ATLAS detector in
  proton-proton collisions at $\sqrt{s} = 7$~TeV}\/},   (2011)  ,
  \href{http://arxiv.org/abs/1112.6426}{{\tt arXiv:1112.6426 [hep-ex]}}.
  accepted by Eur. Phys. J. C.

\bibitem{Collaboration2011}
{ATLAS Collaboration}, {\em {Luminosity Determination in $pp$ Collisions at
  $\sqrt{s}=7$~TeV Using the ATLAS Detector at the LHC}\/},
  \href{http://dx.doi.org/10.1140/epjc/s10052-011-1630-5}{Eur. Phys. J. C {\bf
  71} (2011)  1630}.

\bibitem{Rubin2010}
M.~Rubin, G.~P. Salam, and S.~Sapeta, {\em {Giant QCD K-factors beyond NLO}\/},
   \href{http://arxiv.org/abs/1006.2144}{{\tt arXiv:1006.2144 [hep-ph]}}.

\bibitem{Cacciari:2008gp}
M.~Cacciari, G.~P. Salam, and G.~Soyez, {\em {The anti-k$_t$ jet clustering
  algorithm}\/},
\href{http://dx.doi.org/10.1088/1126-6708/2008/04/063}{JHEP {\bf 04} (2008)
  063}.

\bibitem{Cacciari200657}
M.~Cacciari and G.~P. Salam, {\em {Dispelling the $N^{3}$ myth for the $k_t$
  jet-finder}\/},
  \href{http://dx.doi.org/10.1016/j.physletb.2006.08.037}{Phys. Lett. B {\bf
  641} (2006)  57}.

\bibitem{Lampl:1099735}
W.~Lampl, S.~Laplace, D.~Lelas, P.~Loch, H.~Ma, S.~Menke, S.~Rajagopalan,
  D.~Rousseau, S.~Snyder, and G.~Unal, {\em {Calorimeter Clustering Algorithms:
  Description and Performance}\/},
  \href{http://cdsweb.cern.ch/record/1099735}{ATL-LARG-PUB-2008-002} (2008)  .

\bibitem{alpgen}
M.~L. Mangano et al., {\em {ALPGEN, a generator for hard multiparton processes
  in hadronic collisions}\/},
\href{http://dx.doi.org/10.1088/1126-6708/2003/07/001}{JHEP {\bf 07} (2003)
  001}.

\bibitem{Corcella2001}
G.~Corcella et al., {\em {HERWIG 6: an event generator for hadron emission
  reactions with interfering gluons (including supersymmetric processes)}\/},
  \href{http://dx.doi.org/10.1088/1126-6708/2001/01/010}{JHEP {\bf 01} (2001)
  010}.

\bibitem{jimmy}
J.~M. Butterworth, J.~R. Forshaw, and M.~H. Seymour, {\em {Multiparton
  interactions in photoproduction at HERA}\/},
\href{http://dx.doi.org/10.1007/s002880050286}{Z. Phys. C {\bf 72} (1996)
  637}.

\bibitem{PDF-CTEQ}
J.~Pumplin et al., {\em {New generation of parton distributions with
  uncertainties from global QCD analysis}\/},
\href{http://dx.doi.org/10.1088/1126-6708/2002/07/012}{JHEP {\bf 07} (2002)
  012}.

\bibitem{Sjostrand2001}
T.~Sj\"{o}strand, {\em {High-energy-physics event generation with PYTHIA
  6.1}\/},  \href{http://dx.doi.org/10.1016/S0010-4655(00)00236-8}{Comput.
  Phys. Comm. {\bf 135} (2001)  238}.

\bibitem{Bahr2008}
M.~Bahr et al., {\em {Herwig++ Physics and Manual}\/},
\href{http://dx.doi.org/10.1140/epjc/s10052-008-0798-9}{Eur. Phys. J. C {\bf
  58} (2008)  639}.

\bibitem{Perugia2010}
P.~Z. Skands, {\em {Tuning Monte Carlo Generators: The Perugia Tunes}\/},
\href{http://dx.doi.org/10.1103/PhysRevD.82.074018}{Phys. Rev. D {\bf 82}
  (2010)  074018}.

\bibitem{Martin:2009iq}
A.~D. Martin, W.~J. Stirling, R.~S. Thorne, and G.~Watt, {\em {Parton
  distributions for the LHC}\/},
\href{http://dx.doi.org/10.1140/epjc/s10052-009-1072-5}{Eur. Phys. J. C {\bf
  63} (2009)  189}.

\bibitem{Sherstnev:2007nd}
A.~Sherstnev and R.~S. Thorne, {\em {Parton Distributions for LO
  Generators}\/},
\href{http://dx.doi.org/10.1140/epjc/s10052-008-0610-x}{Eur. Phys. J. C {\bf
  55} (2008)  553}.

\bibitem{Collaboration:2011tq}
{ATLAS Collaboration}, {\em {Measurement of multi-jet cross sections in
  proton-proton collisions at a 7 TeV center-of-mass energy}\/},
  \href{http://dx.doi.org/10.1140/epjc/s10052-011-1763-6}{Eur. Phys. J. C {\bf
  71} (2011)  1763}.

\bibitem{Collaboration:2011kq}
{ATLAS Collaboration}, {\em {Study of jet shapes in inclusive jet production in
  pp collisions at $\sqrt{s} = 7$~TeV using the ATLAS detector}\/},
  \href{http://dx.doi.org/10.1103/PhysRevD.83.052003}{Phys. Rev. D {\bf 83}
  (2011)  052003}.

\bibitem{simulation}
{ATLAS Collaboration}, {\em {The ATLAS Simulation Infrastructure}\/},
  \href{http://dx.doi.org/10.1140/epjc/s10052-010-1429-9}{Eur. Phys. J. C {\bf
  70} (2010)  823}.

\bibitem{geant4}
{GEANT4} Collaboration, S.~Agostinelli et al., {\em {GEANT4: A simulation
  toolkit}\/},
\href{http://dx.doi.org/10.1016/S0168-9002(03)01368-8}{NIM A {\bf 506} (2003)
  250}.

\bibitem{QGS}
G.~Folger and J.~P. Wellisch, {\em {String parton models in Geant4}\/},
\href{http://arxiv.org/abs/nucl-th/0306007}{{\tt arXiv:nucl-th/0306007}}.

\bibitem{Bertini}
H.~W. Bertini, {\em {Intranuclear-Cascade Calculation of the Secondary Nucleon
  Spectra from Nucleon-Nucleus Interactions in the Energy Range 340 to 2900 MeV
  and Comparisons with Experiment}\/},
\href{http://dx.doi.org/10.1103/PhysRev.188.1711}{Phys. Rev. A {\bf 188} (1969)
   1711}.

\bibitem{Fritiof}
B.~Andersson, G.~Gustafson, and B.~Nilsson-Almqvist, {\em A model for
  low-$p_{T}$ hadronic reactions with generalizations to hadron-nucleus and
  nucleus-nucleus collisions\/},
  \href{http://dx.doi.org/10.1016/0550-3213(87)90257-4}{Nucl. Phys. B {\bf 281}
  (1987)  289}.

\bibitem{ATLASWJetspaper2010}
{ATLAS Collaboration}, {\em {Study of jets produced in association with a W
  boson in $pp$ collisions at $\sqrt{s} = 7$~TeV}\/},
  \href{http://dx.doi.org/10.1103/PhysRevD.85.092002}{Phys. Rev. D {\bf 85}
  (2012)  092002}.

\bibitem{ATLASJetFragmentation2010}
{ATLAS Collaboration}, {\em {Measurement of the jet fragmentation function and
  transverse profile in proton-proton collisions at a center-of-mass energy of
  7 TeV with the ATLAS detector}\/},
\href{http://dx.doi.org/10.1140/epjc/s10052-011-1795-y}{Eur. Phys. J. C {\bf
  71} (2011)  1795}.

\end{thebibliography}\endgroup
